\documentclass[aps,pra,twocolumn,amsmath,amssymb,nofootinbib,superscriptaddress]{revtex4-1}

\usepackage{times}
\usepackage[pdftex]{graphicx}
\usepackage{dcolumn}
\usepackage{bm}
\usepackage{amsmath}
\usepackage{indentfirst}
\usepackage{float}
\usepackage[colorlinks]{hyperref}
\usepackage[dvipsnames]{xcolor}
\usepackage{xcolor}
\usepackage{makecell}

\begin{document}

\title{Superconducting Quantum Computing: A Review}

\author{He-Liang Huang}
\email{quanhhl@ustc.edu.cn}
\affiliation{Hefei National Laboratory for Physical Sciences at Microscale and Department of Modern Physics,\\
University of Science and Technology of China, Hefei, Anhui 230026, China}
\affiliation{Shanghai Branch, CAS Centre for Excellence and Synergetic Innovation Centre in Quantum Information and Quantum Physics,\\
University of Science and Technology of China, Hefei, Anhui 201315, China}

\author{Dachao Wu}
\affiliation{Hefei National Laboratory for Physical Sciences at Microscale and Department of Modern Physics,\\
University of Science and Technology of China, Hefei, Anhui 230026, China}
\affiliation{Shanghai Branch, CAS Centre for Excellence and Synergetic Innovation Centre in Quantum Information and Quantum Physics,\\
University of Science and Technology of China, Hefei, Anhui 201315, China}

\author{Daojin Fan}
\affiliation{Hefei National Laboratory for Physical Sciences at Microscale and Department of Modern Physics,\\
University of Science and Technology of China, Hefei, Anhui 230026, China}
\affiliation{Shanghai Branch, CAS Centre for Excellence and Synergetic Innovation Centre in Quantum Information and Quantum Physics,\\
University of Science and Technology of China, Hefei, Anhui 201315, China}

\author{Xiaobo Zhu}
\email{xbzhu16@ustc.edu.cn}
\affiliation{Hefei National Laboratory for Physical Sciences at Microscale and Department of Modern Physics,\\
University of Science and Technology of China, Hefei, Anhui 230026, China}
\affiliation{Shanghai Branch, CAS Centre for Excellence and Synergetic Innovation Centre in Quantum Information and Quantum Physics,\\
University of Science and Technology of China, Hefei, Anhui 201315, China}

\date{\today}


\begin{abstract}
Over the last two decades, tremendous advances have been made for constructing large-scale quantum computers. In particular, the quantum processor architecture based on superconducting qubits has become the leading candidate for scalable quantum computing platform, and the milestone of demonstrating quantum supremacy was first achieved using 53 superconducting qubits in 2019. In this work, we provide a brief review on the experimental efforts towards building a large-scale superconducting quantum computer, including qubit design, quantum control, readout techniques, and the implementations of error correction and quantum algorithms. Besides the state of the art, we finally discuss future perspectives, and which we hope will motivate further research.
\end{abstract}

\maketitle

\section{Introduction}
Quantum computers harness the intrinsic properties of quantum mechanics, which offers the promise of efficiently solving certain problems that are intractable for classical computers~\cite{feynman1999simulating,shor1994algorithms,boixo2018characterizing}. The most impressive example is that in 1994 Peter Shor showed that quantum computers could efficiently factor numbers~\cite{shor1994algorithms}, which poses a serious threat to RSA encryption. Quantum computers would also have an enormous impact on quantum simulation~\cite{georgescu2014quantum}, and may revolutionize the field of machine learning~\cite{biamonte2017quantum}. Thus, the creation of a practical quantum computer would be a revolutionary achievement. The past few years have witnessed the fast development of quantum computing technologies ~\cite{wright2019benchmarking,wang201818,wang2016experimental,huang2018demonstration,huang2017experimental,wang2019boson,gong2019genuine,ye2019propagation,
song2019generation,omran2019generation,zhang2017observation,arute2019quantum}. In particular, now we have moved into the Noisy Intermediate-Scale Quantum (NISQ) era ~\cite{preskill2018quantum}, and one can expect to control a quantum system over 50 qubits~\cite{zhang2017observation,arute2019quantum}.

Quantum computers can be implemented with a variety of quantum systems, such as trapped ions~\cite{leibfried2003quantum,blatt2012quantum}, superconducting qubits ~\cite{krantz2019quantum,kjaergaard2019superconducting,wendin2017quantum,gu2017microwave,you2006superconducting,you2011atomic,kockum2019quantum,nation2012colloquium}, photons~\cite{wang201818,wang2016experimental,huang2018demonstration,huang2017experimental,wang2019boson}, and silicon~\cite{kane1998silicon,he2019two}. In particular, superconducting qubits have emerged as one of the leading candidate for scalable quantum processor architecture. In 1999, a simple qubit for superconducting computing was developed~\cite{nakamura1999coherent}. Subsequently, especially in recent years, superconducting quantum computing has developed rapidly, the number of qubits is rapidly scaling up, and the quality of qubits is also rapidly improving. In 2014, high fidelity (99.4$\%$) two-qubit gate using five qubits superconducting quantum system was achieved~\cite{barends2014superconducting}, which provides an important step toward surface code scheme~\cite{fowler2012surface}. The major milestone, known as quantum supremacy~\cite{boixo2018characterizing}, represents a long-sought stride towards quantum computing, was first demonstrated using superconducting quantum system in 2019~\cite{arute2019quantum}. Due to the rapid development of superconducting quantum computing, the global race to the quantum computer is in full swing. Many technology industries, including Google, IBM, Microsoft, and as well as Intel, are jockeying for a position in quantum computing. All these advances and efforts have brought a promising future for superconducting quantum computing.

Here we provide an overview of superconducting quantum computing, including the basic theoretical ideas, the qubit design, quantum control, readout techniques, and experimental progress in this field. Session 2 discusses different types of superconducting qubits and the development of the lifetime of qubits in recent years. Session 3 and Session 4 introduce the qubit control and qubit readout techniques for superconducting quantum computing. Session 5 and Session 6 review the progress in the experiments for error correction and quantum algorithms. Finally, we briefly summarize our discussions and provide an outlook in Session 7.

\section{Superconducting qubit}

Qubit design has a significant impact on creating devices with high performance, such as long coherence time, high controllability, and so on. In this section, we will briefly introduce several types of superconducting qubits.

\subsection{The DiVincenzo criteria}
As early as 2000, a series of criteria have been summarized to test whether a certain physical system can be used to realize quantum computing. In 2000, David DiVincenzo laid out the first organized set of characteristics our hardware has to exhibit for us to build a quantum computer, which is called ``DiVincenzo criteria" ~\cite{divincenzo2000physical}, they are:

1. A scalable physical system with well characterized qubit.

2. The ability to initialize the state of the qubits to a simple fiducial state.

3. Long relevant decoherence times.

4. A ``universal" set of quantum gates.

5. A qubit-specific measurement capability.

6. The ability to interconvert stationary and flying qubits.

7. The ability to faithfully transmit flying qubits between specified locations.

The first five are known as DiVincenzo's quantum computation criteria, and the last two are necessary for quantum communication. These prerequisites have become the basis for people to find and screen physical systems that may be used to realize quantum computing.


\subsection{Advantages of superconducting qubits}

Superconducting qubits are solid state electrical circuits. Compared with the qubits based on other quantum systems (such as trapped ions, nuclear magnetic resonance, linear optical systems, etc.), superconducting qubits have the following advantages:

1. High designability. Superconducting qubit system has high designability. Different types of qubits, such as charge qubits, flux qubits, and phase qubits, can be designed. And different parameters, such as the energy level of the qubit and the coupling strength, can also be adjusted by adjusting the capacitance, inductance, and Josephson energy. Thus, the Hamiltonian of superconducting qubits can be designed.

2. Scalability. The preparation of superconducting qubits is based on the existing semiconductor microfabrication process. High-quality devices can be prepared by leveraging advanced chip-making technologies, which is good for manufacturing and scalability.

3. Easy to couple. The circuit nature of the superconducting qubit system makes it relatively easy to couple multiple qubits together. In general, superconducting qubits can be coupled by capacitance or inductance.

4. Easy to control. The operation and measurement of superconducting qubits compatible with microwave control and operability. Thus, commercial microwave devices and equipment can be used in superconducting quantum computing experiments.

These advantages make superconducting qubits become the leading candidate for scalable quantum computing. However, to build a large-scale quantum computing, there still remain some outstanding challenges. Due to the tunability and large size of superconducting qubits, the primary disadvantage of superconducting qubits is their short coherence times. Superconducting qubits are not true 2-level systems, thus the unwanted $|1\rangle  \to |2\rangle$ transition must be carefully avoided during information processing. Superconducting qubits require dilution refrigerators to maintain temperatures low enough, and advances of the capacity of such cryostats should be made before building a device with millions of qubits. In the last few years, superconducting qubits have seen drastic improvements in coherence time, operation fidelities, fast and high fidelity qubit readout, and even demonstrations of error correction and quantum algorithms, which we will introduce in the following contents.

\subsection{Three superconducting qubit archetypes}

According to different degrees of freedom, superconducting qubits are mainly divided into three categories: charge qubits~\cite{nakamura1999coherent,bouchiat1998quantum}, flux qubits~\cite{mooij1999josephson}, and phase qubits~\cite{martinis2009superconducting}. We can distinguish these tree types of superconducting qubit according to the ratio $E_J/E_C$, where $E_J$ is Josephson energy and $E_C$ is charging energy.

\begin{figure}[!t]
\centering
\includegraphics[width=0.9\columnwidth]{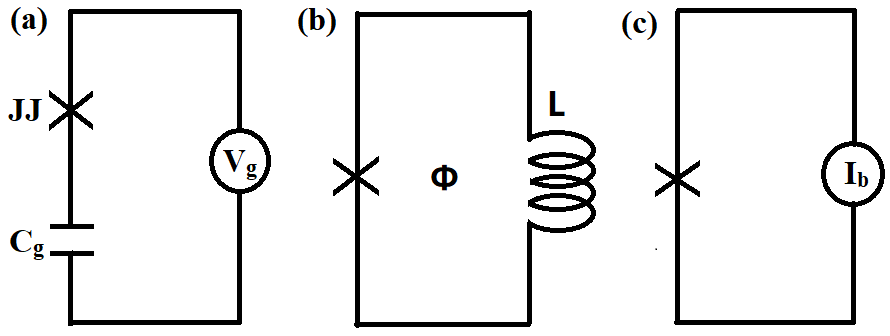}
\caption{Superconducting qubit circuit diagram. (a) Charge qubit composed of a Josephson junction and a capacitor. Adjusting the voltage $V_g$ can control the number of Cooper pairs. (b) Flux qubit. L is the loop inductance. Changing the bias flux $\Phi$ can adjust the energy level structure of the qubit. (c) Phase qubit. Adjusting the bias current $I_b$ can tilt the potential energy surface.}
\label{Figure 2.1}
\end{figure}

The charge qubit is is also called Cooper-pair box qubit, which is shown in Fig.~\ref{Figure 2.1}(a). For the charge qubit, the electrostatic energy of the Cooper-pair on the superconducting island is much larger than the Josephson coupling energy on the junction $E_{J}\ll{E_{C}}$. The relevant quantum variable is the number of Cooper pairs that cross Josephson junction. Fig.~\ref{Figure 2.1}(b) is a schematic diagram of a flux qubit. Typically, $E_J/E_C$ is much lager than 1 but smaller than 100 for flux qubit. The states of the flux qubit are distinguished by the direction of the continuous current state in the superconducting loop. As shown in Fig.~\ref{Figure 2.1}(c), the phase qubit is a current-biased single-junction device with a large ratio $E_J/E_C$ ($E_{J}\gg E_C$). Using the two lowest energy levels of the washboard potential, one phase qubit can be encoded.

\subsection{New types of superconducting qubits}

Based on the three superconducting qubit archetypes, many new types of superconducting qubits are derived, such as Transmon-type qubit, C-shunt flux qubit, Fluxonium, 0-$\pi$qubit, hybrid qubit, and so on. Some are briefly introduced below.

\subsubsection{Transmon-type qubit}

\begin{figure}[!t]
\centering
\includegraphics[width=0.8\columnwidth]{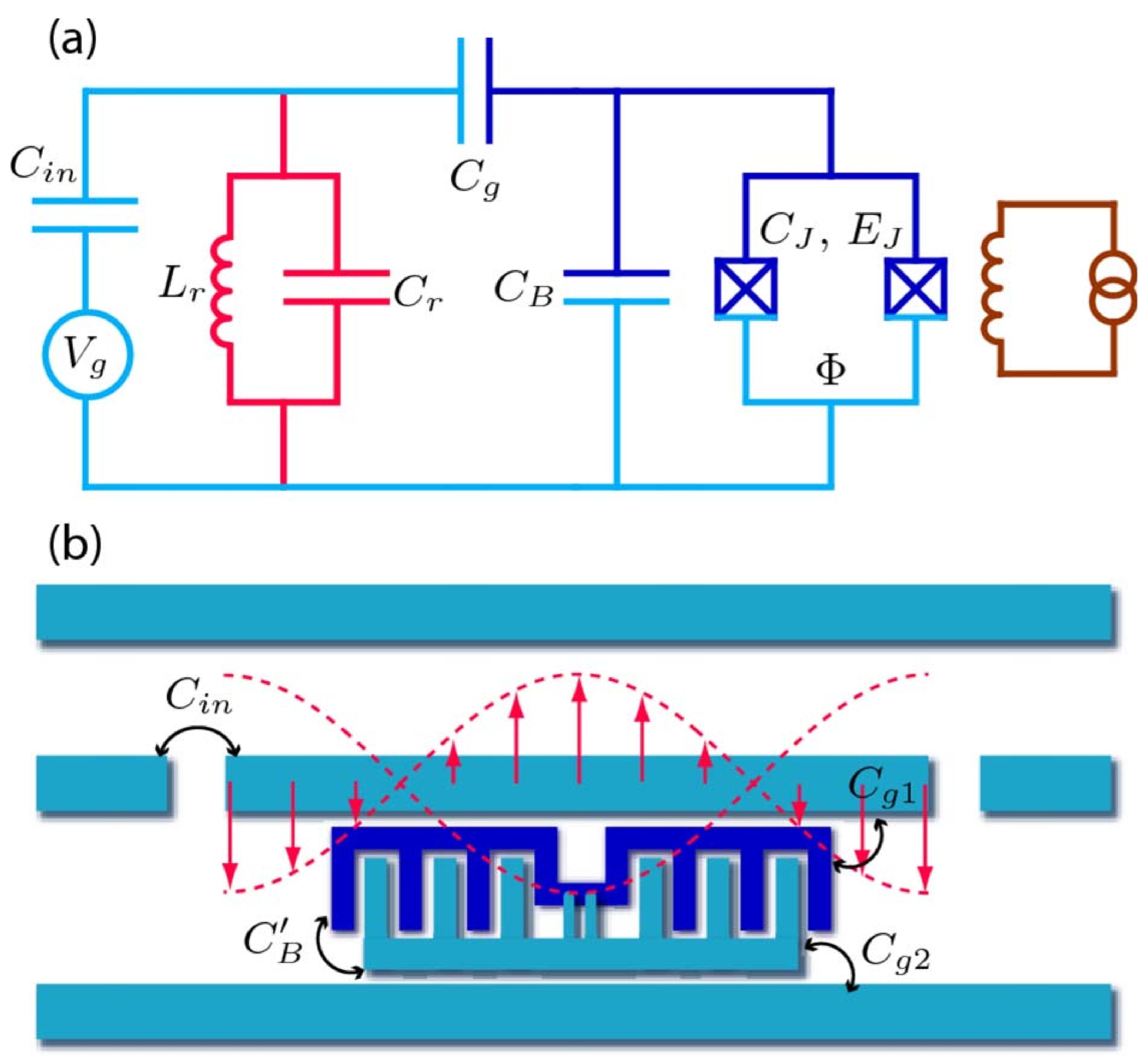}
\caption{Schematic of a Transmon qubit and its effective circuit. (a) The effective circuit model of a transmon qubit. $C_B$ is a large capacitor in parallel with superconducting quantum interference device (SQUID). $L_r$ and $C_r$ are connected in parallel to form an equivalent circuit of the readout resonator. The circuit on the far right is the flux bias of SQUID. (b) Schematic of 2D structure of a transmon qubit. Taken from ~\cite{koch2007charge}.}
\label{fig:Figure 2.2}
\end{figure}

Transmon-type qubit, including Transmon, Xmon, Gmon, 3D Transmon etc., is currently the most popular superconducting qubit due to its simplicity and the flexibility of cQED architectures. Transmon was first proposed by Koch $et$ $al.$ in 2007~\cite{koch2007charge}. In the charge qubit, the charge dispersion decreases exponentially in $E_{J}/E_{C}$, and the anharmonicity algebraically decreases with a slow power law of $E_{J}/E_{C}$. Therefore, properly increasing $E_{J}/E_{C}$ can greatly reduce the sensitivity of the system to charges, while still maintaining sufficient anharmonicity. As shown in Fig.~\ref{fig:Figure 2.2}, the charge energy of Transmon qubit are decreased by paralleling a large interdigitated capacitor across the Josephson junction, thereby obtaining $E_{J}/E_{C}$$\sim$100.

\begin{figure}[!t]
\centering
\includegraphics[width=0.9\columnwidth]{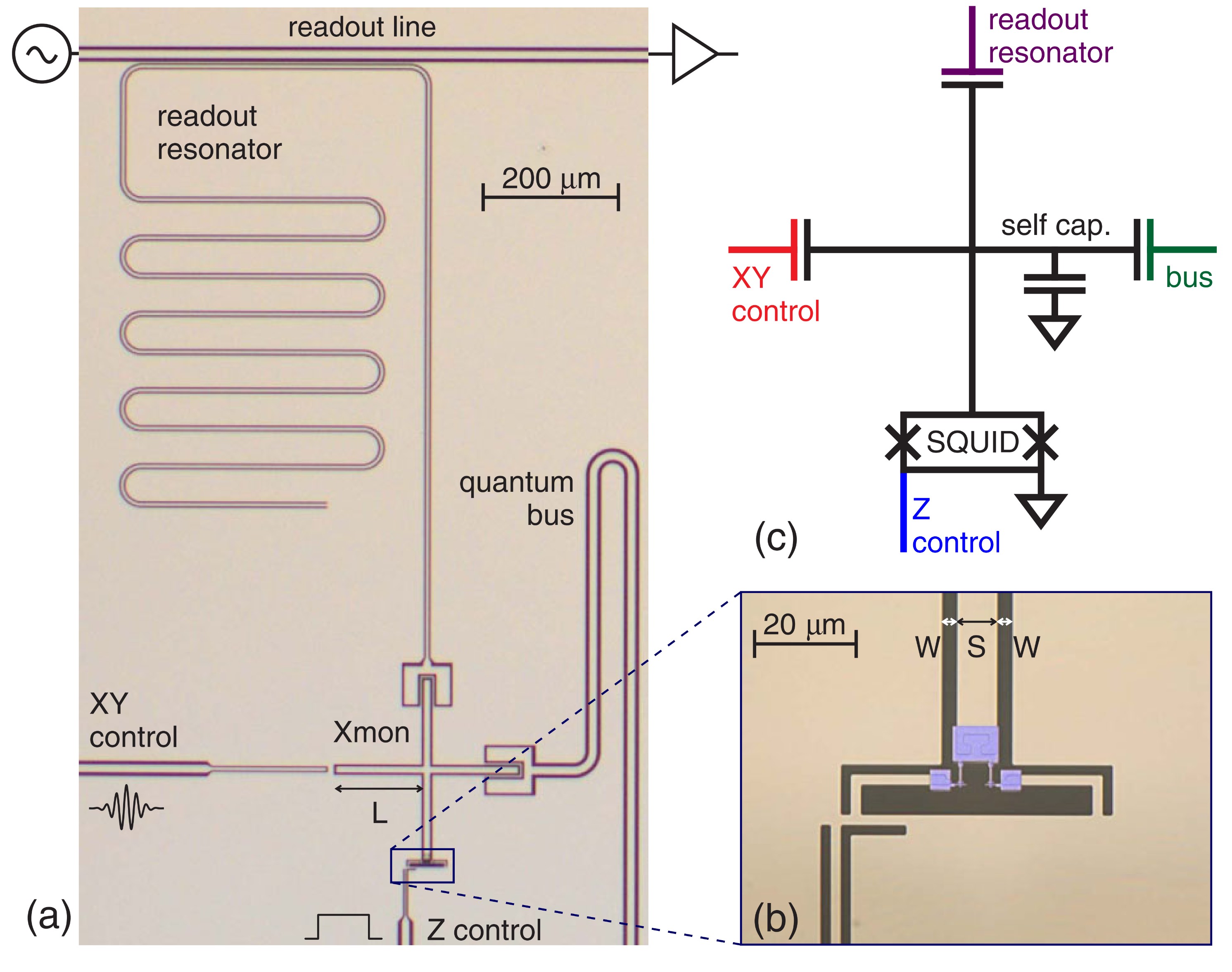}
\caption{(a) Optical micrograph of Xmon qubit. (b) Enlarged image of SQUID. (c) The electrical circuit of the qubit. Taken from~\cite{barends2013coherent}.}
\label{fig:Figure 2.3}
\end{figure}

Xmon can be considered as a modified version of Transmon, which was proposed by Barends $et$ $al.$ in 2013~\cite{barends2013coherent}. As shown in Fig.~\ref{fig:Figure 2.3}, different from the original Transmon's interdigital capacitor, the Xmon qubit is formed by a cross capacitor, and the Xmon qubit is coupled to a common transmission line through a resonant cavity. Each Xmon is controlled by two independent control lines, a $XY$ control line and a $Z$ control line, which can be used for rotating the quantum state in the $X$, $Y$ and $Z$ directions. The qubits can be directly coupled through capacitance. The Xmon qubit combines fast control, long coherence, and straightforward connectivity, which is suitable for scalable superconducting quantum computing.

\begin{figure}[!t]
\centering
\includegraphics[width=0.8\columnwidth]{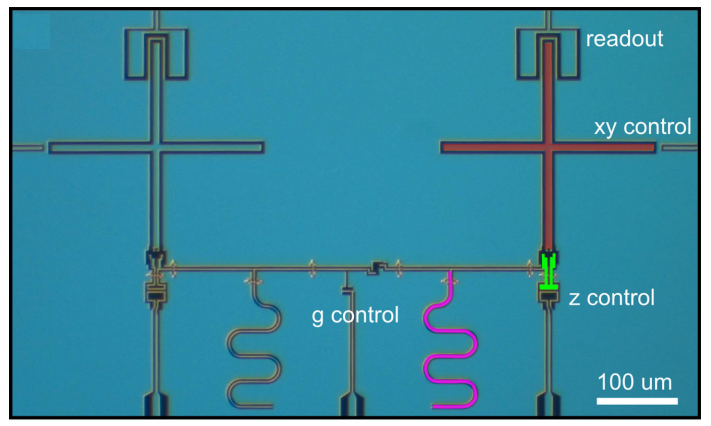}
\caption{Optical micrograph of two inductively coupled Gmon qubits. Taken from~\cite{chen2014qubit}.}
\label{fig:Figure 2.4}
\end{figure}

\begin{figure}[!t]
\centering
\includegraphics[width=0.9\columnwidth]{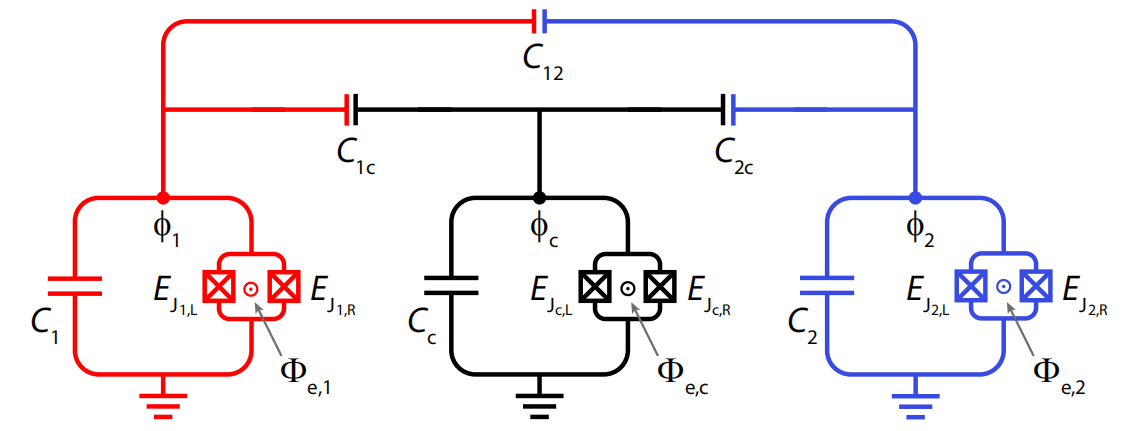}
\caption{Circuit diagram of a superconducting circuit implementing a tunable coupler, consisting of two qubit modes (red and blue) and a coupler mode (black). Taken from~\cite{yan2018tunable}.} 
\label{fig:Figure 2.5}
\end{figure}

Gmon~\cite{chen2014qubit} is based on the Xmon qubit, but the qubits are connected with a junction serving as a tunable inductor to control the coupling strength (see Fig.~\ref{fig:Figure 2.4}). The Gmon architecture is protected from the problem of frequency crowding that arise from fixed coupling, and provides a flexible
platform with applications ranging from quantum computing to quantum simulation. However, the extra inductive coupler introduces additional decoherent channels, and the device layout becomes very complicated. In 2018, Yan $et$ $al.$ proposed a simple and generic scheme for a tunable coupler~\cite{yan2018tunable}, which is a generic three-body system in a chain geometry, and the center mode is a tunable coupler (see Fig.~\ref{fig:Figure 2.5}). The center mode can be constructed with any flux-tunable circuit in which the resonance frequency can be tuned. By modulating the coupler frequency, the coupling strength of the two next-nearest neighbor qubits can be tuned. This architecture does not introduce additional components, and its layout is similar to the Xmon.

\begin{figure}[!t]
\centering
\includegraphics[width=0.9\columnwidth]{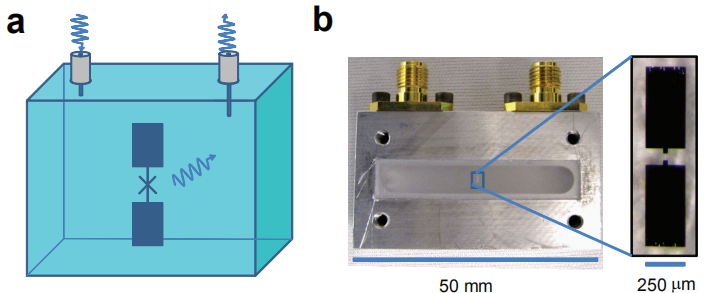}
\caption{(a) Schematic of a transmon qubit inside a 3D cavity. (b) Photograph of a half of the 3D aluminum waveguide cavity. Taken from~\cite{paik2011observation}.}
\label{fig:Figure 2.6}
\end{figure}

3D Transmon is a development of Transmon qubits, which was proposed by Paik $et$ $al.$ in 2011~\cite{paik2011observation}. The biggest feature of 3D Transmon is the replacement of the planar transmission-line cavities with a three-dimensional waveguide cavity (see Fig.~\ref{fig:Figure 2.6}), which offers several advantages. First, the cavity has a larger mode volume and is much less sensitive to the surface dielectric losses. Second, the architecture provides the qubit with a well-controlled electromagnetic environment. Thus, this architecture could suppress the decoherence of qubit while maintaining sufficient coupling to the control signal. However, scalability is the major difficulty we have to overcome, if we want to build a large-scale device based on 3D Transmon.

\subsubsection{3-JJ flux qubit}

\begin{figure}[!t]
\centering
\includegraphics[width=0.5\columnwidth]{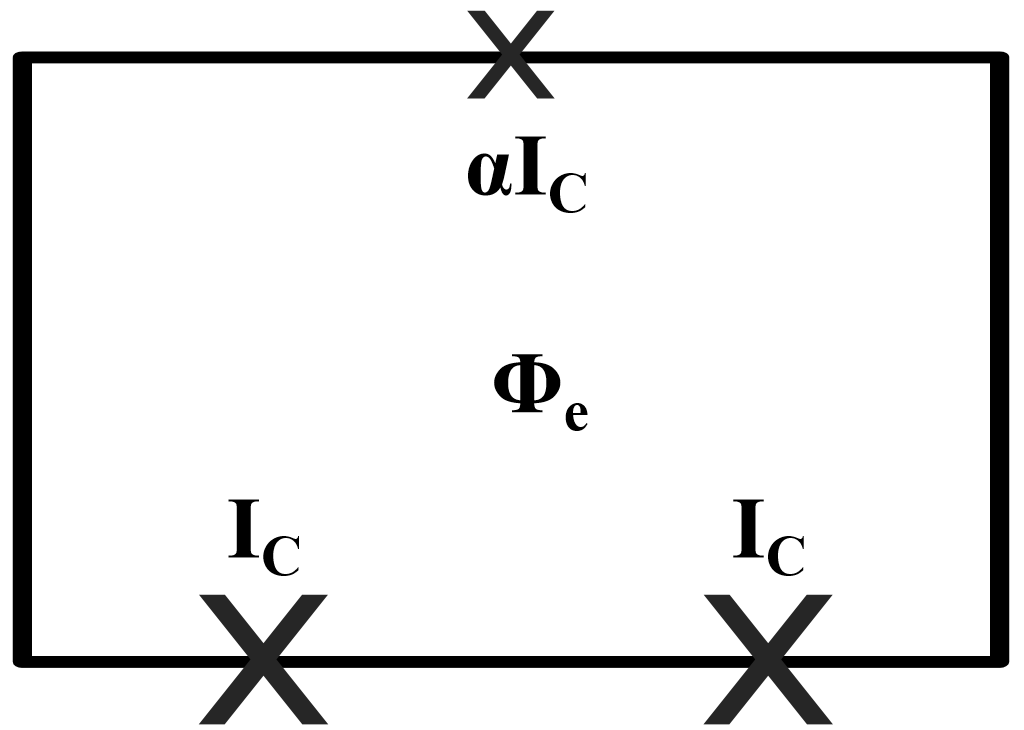}
\caption{Schematic diagram of 3-JJ flux qubit. Two junctions have the same Josephson coupling energy $E_J$, and the third junction has a smaller Josephson coupling energy ${\alpha}E_J$.}
\label{fig:Figure 2.8}
\end{figure}

The three-Josephson-junction (3-JJ) flux qubit was proposed by Mooij $et$ $al.$ in 1999~\cite{mooij1999josephson}, which consists of a micrometer-sized loop with three or four Josephson junctions (see Fig.~\ref{fig:Figure 2.8}). The reduction in the size of the loop in 3-JJ flux qubit, resulting a reduced sensitivity of the magnetic flux qubit to magnetic flux noise. In this architecture, the two qubit states have persistent current in opposite directions, and the quantum superposition of these two quantum states could be obtained by pulsed microwave modulation of the enclosed magnetic flux by current in control lines.

\subsubsection{C-shunt flux qubit}

\begin{figure}[!t]
\centering
\includegraphics[width=0.8\columnwidth]{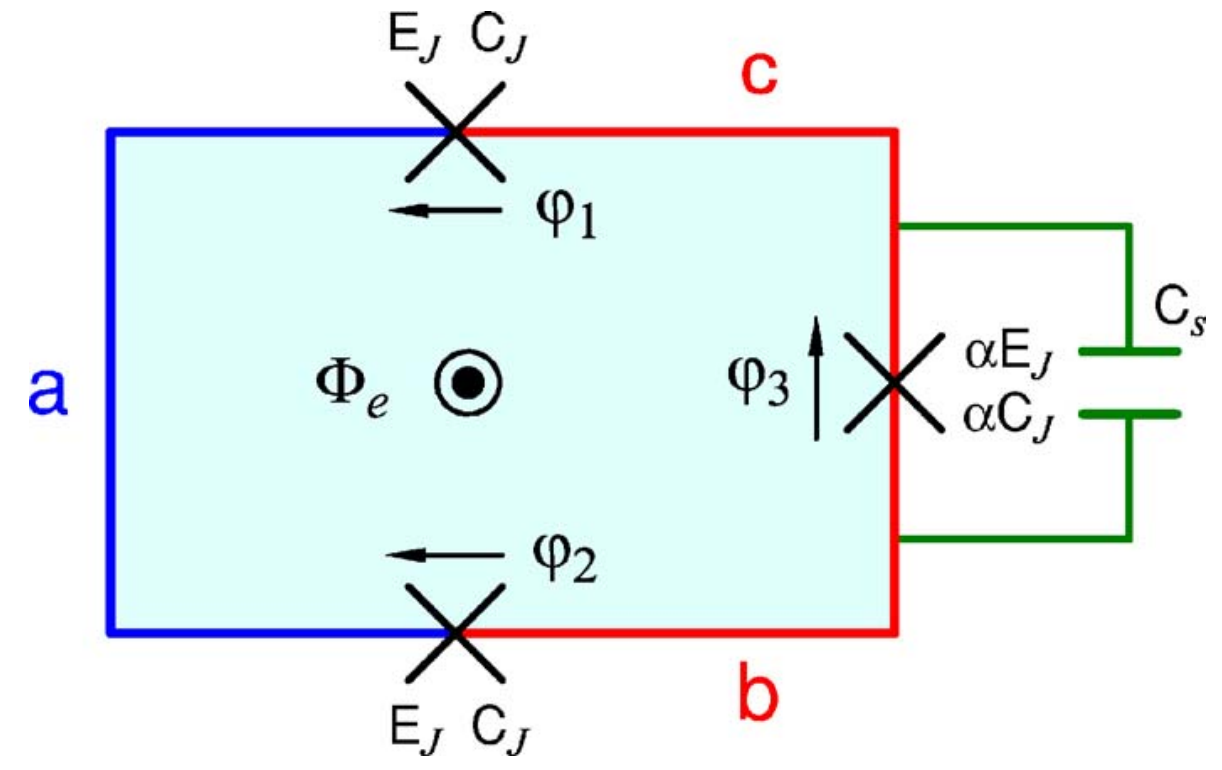}
\caption{Schematic diagram of C-shunt flux qubit. A capacitance $C_s$ is shunted in parallel to the smaller Josephson junctions to reduce the charging energy related to islands b and c. Taken from~\cite{you2007low}.}
\label{fig:Figure 2.9}
\end{figure}

The capacitively shunted (C-shunt) flux qubit was proposed by You $et$ $al.$ in 2007~\cite{you2007low}. In the 3-JJ flux qubit, the effect of flux noise has been greatly suppressed, and the charge noise is the main source of decoherence, which mainly comes from the charge fluctuations on the two islands separated by the smaller Josephson junction. Compared with 3-JJ flux qubits, C-shunt flux qubit architecture introduces an additional capacitor shunted in parallel to the smaller Josephson junction in the loop (see Fig.~\ref{fig:Figure 2.9}). This shunt capacitance is used to reduce the charging energy, thus the effects of the dominant charge noise in the two islands are suppressed. This architecture could suppress the effects of both flux and charge noises.

\subsubsection{Fluxonium}

\begin{figure}[!t]
\centering
\includegraphics[width=0.7\columnwidth]{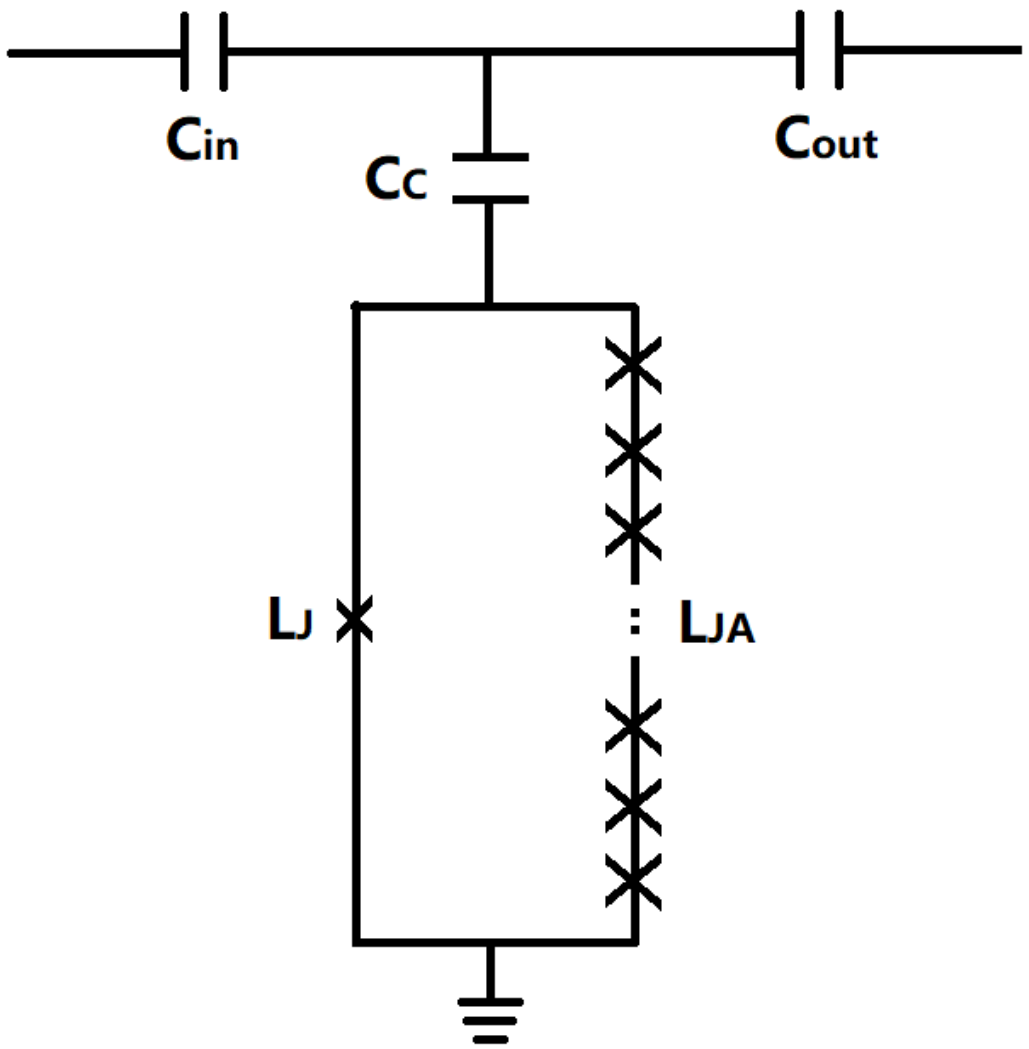}
\caption{The electrical circuit representation of Fluxonium qubit.}
\label{fig:Figure 2.10}
\end{figure}

Fluxonium was proposed by Manucharyan $et$ $al.$ in 2009~\cite{manucharyan2009fluxonium} to solves both the inductance and the offset charge noise problems. In the Fluxonium architecture, a series array of large-capacitance tunnel junctions are connected in parallel with a small junction (see Fig.~\ref{fig:Figure 2.10}). When the system oscillation frequency is below the plasma frequency, the series array of large junctions effectively behaves as an inductive wire. From a circuit perspective, such a large inductor is equivalent to a low-pass filter. Therefore, the low-frequency change of the charge across the small junction is short-circuited by this large inductor, which reduces the sensitivity of qubits to charge noise.

\subsubsection{0-$\displaystyle \pi$ qubit}

\begin{figure}[!t]
\centering
\includegraphics[width=\columnwidth]{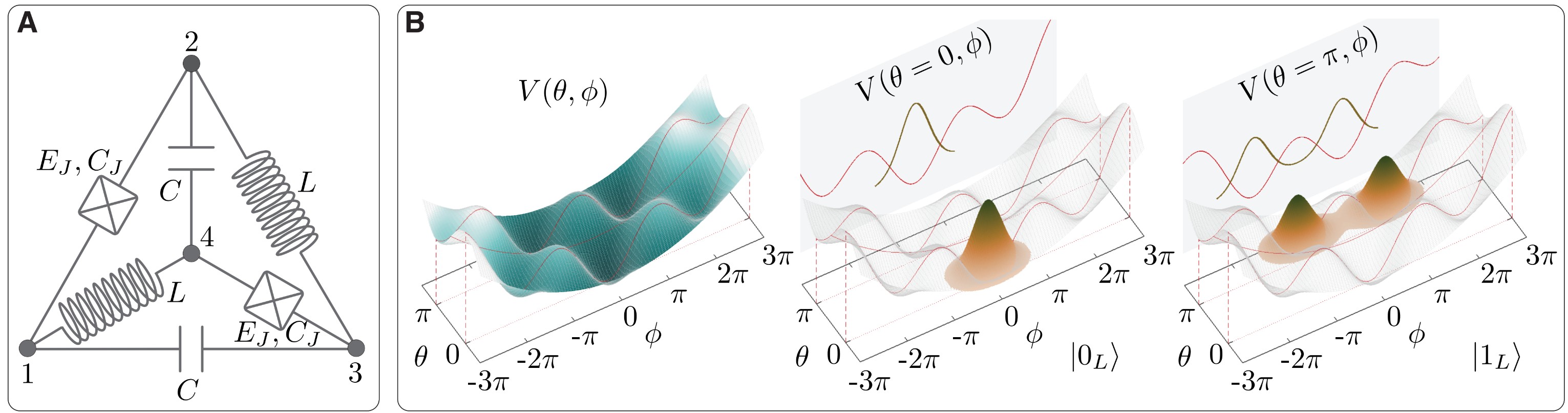}
\caption{(a) Circuit diagram of 0-$\displaystyle \pi$ superconducting qubits. The circuit has a ring with four nodes. The four nodes are connected by a pair of Josephson junctions ($E_J$, $C_J$), a large capacitor ($C$) and superinductors ($L$). (b) In the absence of a magnetic field, the double-well potential function $V(\theta ,\phi )$ of the circuit. The ground state of the 0 valley is localized along $\theta=0$, and the lowest state of the $\displaystyle \pi$ valley is localized along $\theta=\displaystyle \pi$. Taken
from~\cite{gyenis2019experimental}.}
\label{fig:Figure 2.11}
\end{figure}

The 0-$\displaystyle \pi$ qubit was first proposed by Kitaev, Brooks and Preskill~\cite{kitaev2006protected, brooks2013protected} and experimentally realized by Gyenis $et$ $al.$~\cite{gyenis2019experimental} in 2019. 0-$\displaystyle \pi$ qubit is designed with a symmetrical circuit to obtain an interleaved double potential well (see Fig.\ref{fig:Figure 2.11}). The two ground state wave functions of a qubit are highly localized in their respective potential wells and do not disjoint each other. The transition matrix elements between the corresponding two ground state energy levels are very small. Therefore, 0-$\displaystyle \pi$ qubit is not sensitive to charge and magnetic flux noise.

\subsubsection{Hybrid qubit}
\begin{figure}[!t]
\centering
\includegraphics[width=0.75\columnwidth]{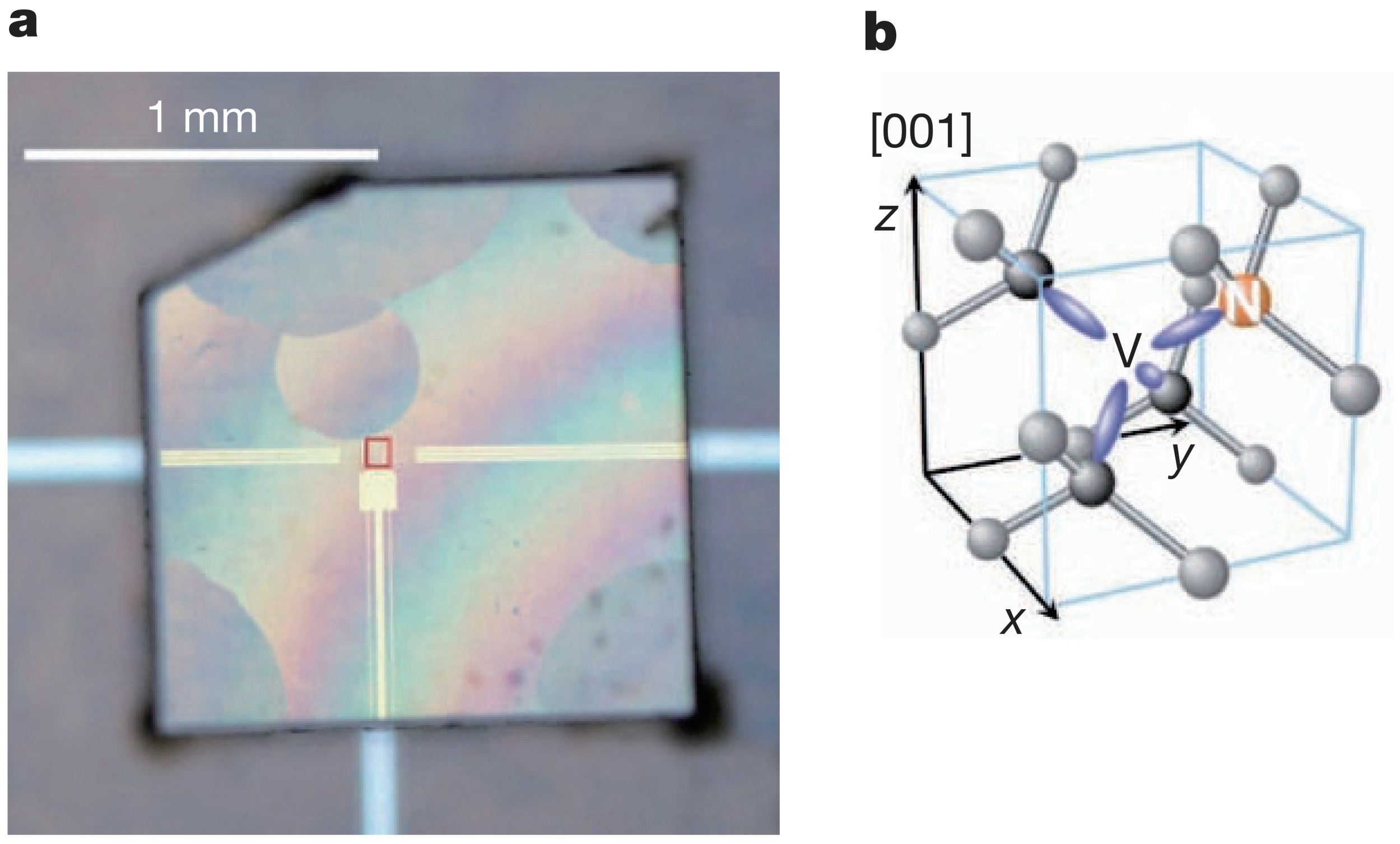}
\caption{Experimental setup of the hybrid system that coupling a superconducting flux qubit to an electron spin ensemble in diamond. (a). Diamond crystal glued on top of a flux qubit (red box). (b). NV centre. Taken from~\cite{zhu2011coherent} .}
\label{fig:Figure 2.7}
\end{figure}

Different quantum systems have their own advantages, and thus hybrid system ~\cite{marcos2010coupling, zhu2011coherent, kubo2011hybrid, schuster2010high, kubo2010strong, amsuss2011cavity, xiang2013hybrid} is proposed to combine the advantages of different quantum systems. In 2010, Marcos $et$ $al.$ proposed a a novel hybrid system that coupling Nitrogen-Vacancy (NV) centers in diamond to superconducting flux qubits ~\cite{marcos2010coupling}. The hybrid system takes advantage of these two systems. The flux qubits are well-controlled, but their coherence time is short, which could be used as a control element. The NV centers have long coherence times, which have the potential to be used as a long-term memory for a superconducting quantum processor. Zhu $et$ $al.$ in 2011 reported the observation of vacuum Rabi oscillations between a flux qubit and an ensemble of approximately 3$\displaystyle \times 10^7$ NV-centers in diamond ~\cite{zhu2011coherent} (see Fig.~\ref{fig:Figure 2.7}). This demonstrates strong coherent coupling between two dissimilar quantum systems with a collective coupling constant of $\displaystyle g_{ens}\approx$70 MHz.

\subsection{Qubit's lifetime}

By improving the structure and parameters of the superconducting qubits, as well as the preparation techniques and materials, the lifetime of the superconducting qubits has been greatly enhanced. TABLE \ref{tbl:table2-1} summarizes the development in the lifetime of qubits over the past decade.

\begin{table}[!htbp]
\centering
\caption{The development in the lifetime of qubits in recent years (each data uses the highest value reported in the literature of the current year, and is not recorded if it is smaller than in previous years.).}
\label{tbl:table2-1}
\begin{tabular}{|c|c|c|c|c|c|c|c|}
\hline
\makecell{$T_1/\mu s$\\ \\} & \makecell{Transmon\\} & \makecell{Xmon} & \makecell{3D \\Transmon}&\makecell{3-JJ \\flux \\qubit}&\makecell{C-\\shunt \\flux \\qubit}&\makecell{ Flux-\\onium}& \makecell{0-$\pi$\\qubit} \\
\hline
2010& 1.2~\cite{dicarlo2010preparation}& & &4~\cite{fedorov2010strong}&1.5~\cite{steffen2010high}& & \\
\hline
2011& 1.6~\cite{hoffman2011coherent} & & 60~\cite{paik2011observation} &12~\cite{bylander2011noise}&5.7~\cite{corcoles2011protecting}& & \\
\hline
2012& 9.7~\cite{chow2012universal}  & &70~\cite{rigetti2012superconducting} & & &4~\cite{manucharyan2012evidence} & \\
\hline
2013& 11.6~\cite{corcoles2013process} &44~\cite{barends2013coherent} & & & & & \\
\hline
2014&29~\cite{chow2014implementing} & &95~\cite{wang2014measurement} & & &8100~\cite{pop2014coherent}  & \\
\hline
2015& 36~\cite{corcoles2015demonstration} & & & & & & \\
\hline
2016&56~\cite{takita2016demonstration} & &162~\cite{dial2016bulk} & &55~\cite{yan2016flux} & & \\
\hline
2017& 80~\cite{riste2017demonstration} & & & & & & \\
\hline
2018&  & & & & & & \\
\hline
2019& & & 240~\cite{tsioutsios2019free} & & & &1560~\cite{gyenis2019experimental} \\
\hline
\end{tabular}
\end{table}

\subsection{Number of entangled qubits}

\begin{figure}[!t]
\centering
\includegraphics[width=1\columnwidth]{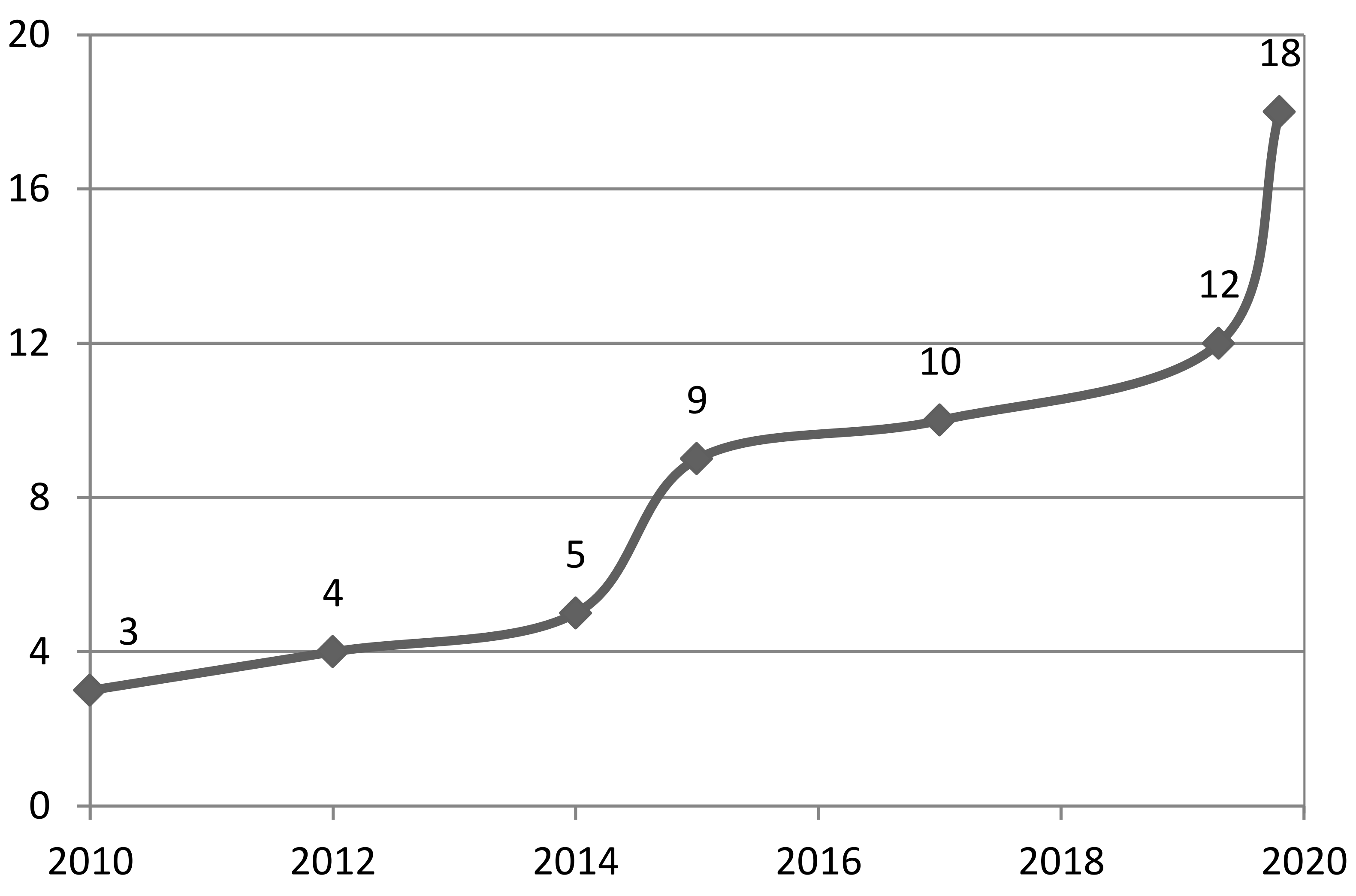}
\caption{The changes in the number of entangled superconducting qubits over the past decade (~\cite{dicarlo2010preparation,lucero2012computing,barends2014superconducting,kelly2015state,song201710,gong2019genuine,song2019generation}).}
\label{fig:FIG2-12}
\end{figure}

The number of superconducting qubits has grown rapidly in recent years. In 2019, the demonstration of  quantum supremacy is first achieved using 53 superconducting qubits ~\cite{arute2019quantum}. The performance of a quantum computer depends on the number and quality of the qubits. The ability to prepare multi-qubit entangled qubits is a critical indicator to show the full qubit control of quantum computing platforms. Here we summarize the changes in the number of entangled superconducting qubits over the past decade (see Fig.~\ref{fig:FIG2-12}).

\section{Qubit Control}

In this section, we will introduce how superconducting qubits are manipulated to implement quantum gate, and the progress of high-fidelity gates.

\subsection{Single-qubit operations}

An arbitrary single-qubit rotation can be defined as

\begin{eqnarray}
R_{\vec{n}}\left( \theta \right) =e^{-i\theta \vec{n}\cdot \vec{\sigma}/2}=\cos \left( \frac{\theta}{2} \right)I -i\sin
\left( \frac{\theta}{2} \right)\left( \vec{n}\cdot \vec{\sigma} \right)
\label{eq3-1}
\end{eqnarray}
where $\vec{n}$ is a real three-dimensional unit vector, and $\vec{\sigma}$ is Pauli matrix. The Single-qubit rotation operations in superconducting quantum circuit can be divided into $XY$ operation and $Z$ operation.

\subsubsection{$XY$ operating principle }

Taking a single Xmon qubit as an example, we usually couple microwave sources to Xmon by capacitance. The microwave drive can be expressed as
$\Omega \left( t \right) =\Omega _x\cos \left( \omega _dt-\phi \right)$, and the driving Hamiltonian can be simply expressed as
\begin{eqnarray}
H=-\frac{\hbar}{2}\omega\sigma_z + \Omega _x\cos \left(\omega _dt-\phi \right) \sigma _x
\label{eq3-2}
\end{eqnarray}
the first term is Xmon's Hamiltonian and the second term is drive term. After transforming the Hamiltonian into the rotating frame, we get
\begin{eqnarray}
H=-\frac{\hbar}{2}\Delta\sigma_z + \frac{\hbar}{2}\Omega _x(\cos\phi\sigma_x+\sin\phi\sigma_y)
\label{eq3-3}
\end{eqnarray}
where $\Delta=\omega-\omega_d $ is the detuning between qubit frequency and microwave frequency. When the qubit resonates with the microwave which means $\Delta=0 $, the first term will be removed, and the angle of the rotation axis in the $XY$-plane determined by the phase $\phi$ of the microwave drive.


\subsubsection{Physical $Z$ operation}

\begin{figure}[!t]
\centering
\includegraphics[width=0.6\columnwidth]{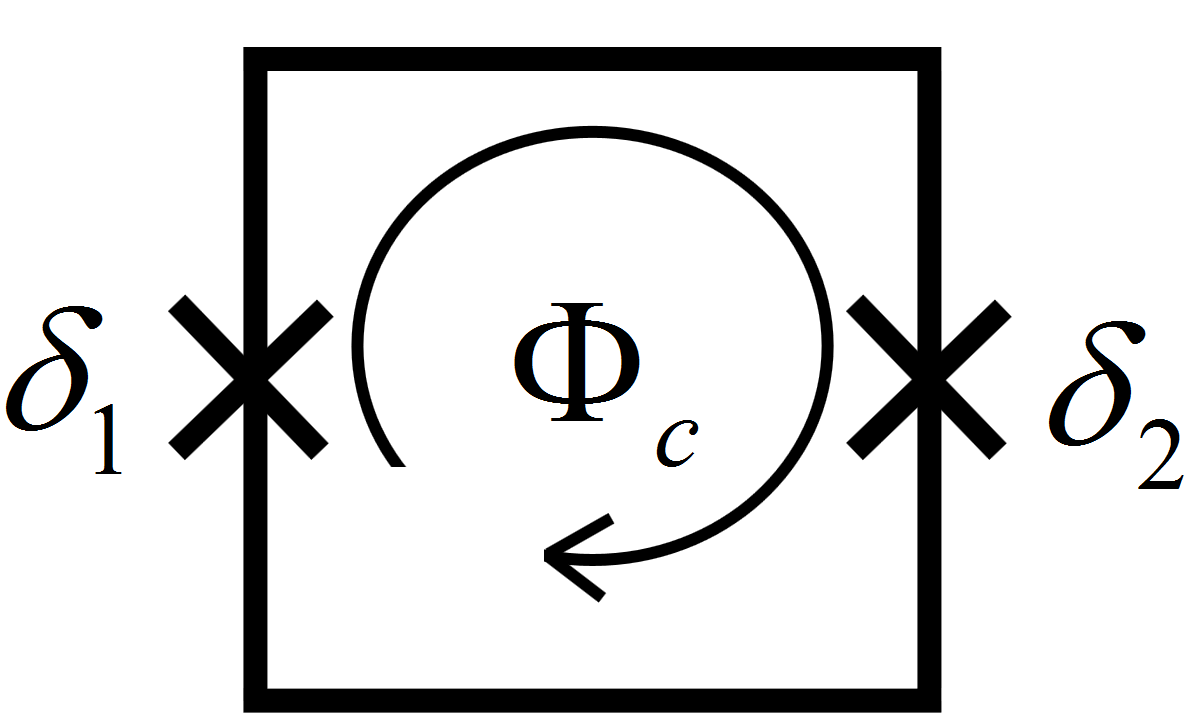}
\caption{Diagram of SQUID loop. ${\delta _1}({\delta _2})$ represent the phase different of Josephson junction. Its critical current varies with the change of extra magnetic flux change.}
\label{fig:Figure 3-1}
\end{figure}

In order to implement physical $Z$ operation, the simplest way is using SQUID loop. The SQUID loop is consisted of two Josephson junctions as shown in the Fig.~\ref{fig:Figure 3-1}. At the bottom of Xmon, $Z$ line will directly connect with SQUID loop. When current flows into SUQID loop by $Z$ line, the loop would generate extra flux, which will cause the frequency of qubit to change. Assuming the extra flux is $\phi_e $, the frequency of qubit would be

\begin{eqnarray}
\omega_q/2\pi=E_1-E_0=\sqrt{8E_cE_J\cos(\frac{\pi\phi_e}{\phi_0})}
\label{eq3-4}
\end{eqnarray}
Then the Hamiltonian will be express as $H=-\frac{\hbar}{2}\omega_q\sigma_z$, and the corresponding evolution operator is
\begin{eqnarray}
U=e^{-\frac i2\int_0^{t_0}-\omega_q(t)\sigma_zdt}=R_z(\int_0^{t_0}-\omega_q(t)dt)
\label{eq3-5}
\end{eqnarray}

Thus, we can rotate the state around $X$-axis in any angel by changing only the magnitude and duration of the current.

\subsubsection{Virtual $Z$ operation}

Virtual $Z$ operation is another and also the most commonly used approach for implementing $Z$ operation, which is realized by just adding a phase offset to the drive field for all subsequent $X$ and $Y$ gates~\cite{MckayUniversal,krantz2019quantum}. We note that the virtual $Z$ operation does not take any time in the experiment, since only additional phases need to be added to the microwave pulse sequence. Thus, it would be very efficient to construct arbitrary single-qubit gates using virtual $Z$ operation. Any single-qubit gate can be written as~\cite{MckayUniversal,krantz2019quantum}

\begin{eqnarray}
U(\theta,\phi,\lambda)= Z_{\phi}\cdot X_{\theta}\cdot Z_{\lambda}
\label{eq3-7}
\end{eqnarray}

for appropriate choice of angles $\theta,\phi,\lambda$. Taking the Hadamard gate an example, the Hadamard gate can be written as $H=Z_{\pi/2}X_{\pi/2}Z_{\pi/2}$. Since the $Z$ operation can be performed with a virtual $Z$ operation, we can implement the Hadamard gate by using only single microwave pulse.

\subsection{Two-qubit gates}


To implement two-qubit gates, the Hamiltonian must have coupling term which is usually $\sigma_x\sigma_x$ and $\sigma_y\sigma_y$ in the superconducting system. For Xmon qubit, if we tune two qubits to the same frequency, the coupling term could be expressed as $H_{couple}=\hbar g(\sigma_1^+\sigma_2^-+\sigma_1^-\sigma_2^+)$ using rotating wave approximation (RWA), and the evolution operator will be

\begin{eqnarray}
U={\rm{exp}}(-\frac i{\hbar}\int_0^{t_0}H_{couple}dt)={\rm{exp}}(-\frac i{\hbar}H_{couple}t)
\label{eq3-6}
\end{eqnarray}

If we let $t=\pi/2g$, then we can get iSWAP gate which has the following matrix form

\begin{eqnarray}
i\rm{SWAP}=\left( \begin{matrix}
	1&		0&		0&		0\\
	0&		0&		i&		0\\
	0&		i&		0&		0\\
	0&		0&		0&		1\\
\end{matrix} \right)
\label{eq7}
\end{eqnarray}

If we cut the interaction time in half, then we would get another useful two-qubit gate, called sqrt(iSWAP) gate

\begin{eqnarray}
\sqrt{i\rm{SWAP}}=\left( \begin{matrix}
	1&		0&		0&		0\\
	0&		1/\sqrt{2}&		i/\sqrt{2}&		0\\
	0&		i/\sqrt{2}&		1/\sqrt{2}&		0\\
	0&		0&		0&		1\\
\end{matrix} \right)
\label{eq8}
\end{eqnarray}

Controlled-not (CNOT) gate and controlled-Z (CZ) gate are two commonly used two-qubit gates, and their matrix form are

\begin{eqnarray}
\rm{CNOT}=\left( \begin{matrix}
	1&		0&		0&		0\\
	0&		1&		0&		0\\
	0&		0&		0&		1\\
	0&		0&		1&		0\\
\end{matrix} \right)
\label{eq9}
\end{eqnarray}

\begin{eqnarray}
\rm{CZ}=\left( \begin{matrix}
	1&		0&		0&		0\\
	0&		1&		0&		0\\
	0&		0&		1&		0\\
	0&		0&		0&		-1\\
\end{matrix} \right)
\label{eq10}
\end{eqnarray}

Next, we will introduce how to implement these two-qubit gates in experiments.

\subsection{Two-qubit implementations}

\subsubsection{Frequency tuning gates}

\begin{figure}[!t]
\centering
\includegraphics[width=0.95\columnwidth]{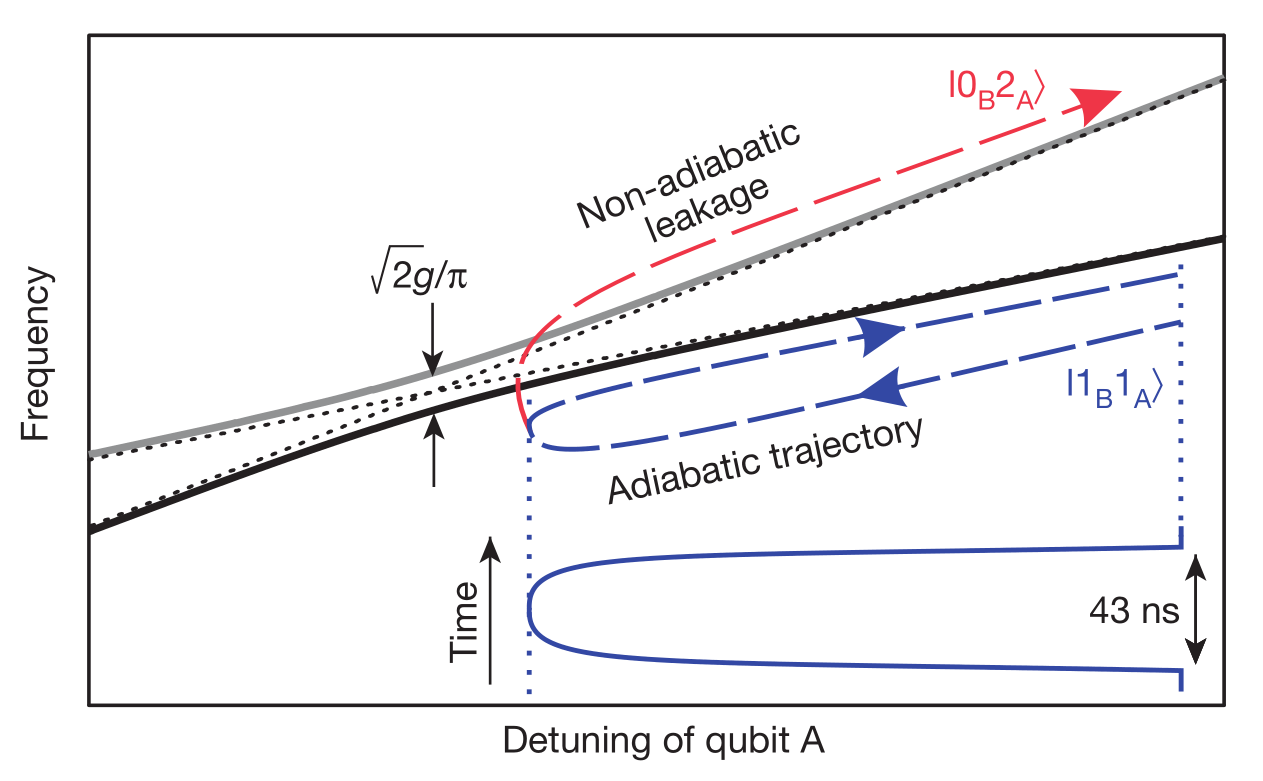}
\caption{Schematic diagram of realizing CZ gate by using the avoided level crossing between $|1,1\rangle$ state and $|0,2\rangle$ state with the fast adiabatic tuning. Taken from~\cite{barends2014superconducting}.}
\label{fig:Figure 3-2}
\end{figure}

A method to implement CZ gate by fast adiabatic tuning is shown in Fig.~\ref{fig:Figure 3-2}, which takes the two-qubit state $|1,1\rangle$ close to the avoided level crossing with the state $|0,2\rangle$, yielding a selective $\pi$  phase change of $|1,1\rangle$ state. In 2009, DiCarlo $et$ $al.$~\cite{dicarlo2009demonstration} demonstrated the implementation of the Grover search and Deutsch$-$Jozsa quantum algorithm using adiabatic controlled-phase gates, and the algorithms are implemented with  fidelity greater than $80\%$. Barends $et$ $al.$~\cite{barends2014superconducting} realized a two-qubit gate by adiabatic tuning with fidelity of $99.4\%$, and the fidelity is characterized by randomized benchmarking~(RB). The high fidelity is achieved by optimizing the pulse amplitude and frequency, and minimizing two-state leakage.

Actually, diabatic tuning could also be used to implement two-qubit gate. Li $et$ $al.$~\cite{li2019realisation} recently designed and implemented non-adiabatic CZ gate, which outperform adiabatic gates in terms of speed and fidelity, with gate times reaching 40 ns, and fidelities reaching $F=99.54\pm0.08\%.$ Barends $et$ $al.$ \cite{barends2019diabatic} reported diabatic two-qubit gates, the iSWAP-like and CPHASE gates, with average gate fidelities of 0.9966(2) and 0.9954(2) respectively by testing using cross-entropy benchmarking (XEB). The gate time is as fast as 18 ns.

\subsubsection{Cross resonance}

\begin{figure}[!t]
\centering
\includegraphics[width=0.8\columnwidth]{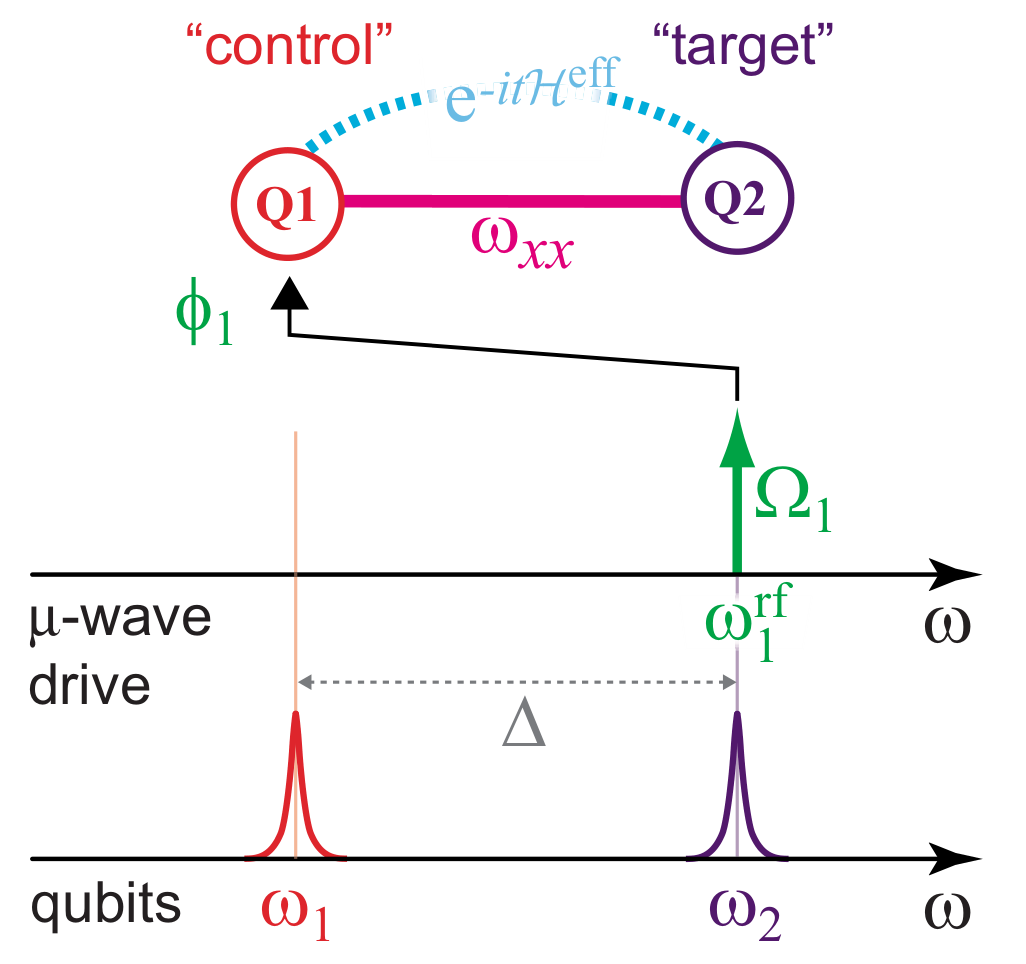}
\caption{FLICFORQ-style qubits (circles) have fixed transition frequencies and fixed linear off-diagonal coupling.  ${\Omega _1}$ is the amplitude of microwave. Q1 and Q2 would weakly couple when $\omega_1^{rf}$ approximately equal to $\Delta$. Taken from~\cite{rigetti2010fully}.}
\label{fig:Figure 3-3}
\end{figure}

The cross-resonance gate is an entangling gate for fixed frequency superconducting qubits, which is implemented by applying a microwave drive to a system
of two coupled qubits as showed Fig.~\ref{fig:Figure 3-3}~\cite{rigetti2010fully,chow2011simple,chow2012universal,chow2014implementing}. The cross-resonance gate was first implemented in 2011 by Chow $et$ $al$.~\cite{chow2011simple}, and they used quantum process tomography (QPT) to reveal a gate fidelity of $81\%$. Corcoles $et$ $al.$~\cite{corcoles2013process} presented a complete RB characterization of two fixed-frequency superconducting qubits, and achieved a gate fidelity of 93.47$\%$ under interleaved RB experiment which compares favorably to the fidelity of 87.99$\%$ obtained by QPT performed on the same gate. Sheldon $et$ $al.$~\cite{sheldon2016procedure} presented improvements implementation of the cross resonance gate with shorter gate time (160 ns) and interleaved RB fidelities exceeding 99$\%$.



\subsubsection{Parametric gates}

The parametric gates are techniques for realizing two-qubit gates using parametric modulation~\cite{beaudoin2012first,strand2013first,mckay2016universal,didier2018analytical,hong2020demonstration,chu2019realization,reagor2018demonstration}. For example, we can mix DC and AC dirve to modulate the flux bias of the two-qubit system to drive population between the $|1,1\rangle$ and $|0,2\rangle$ states. As population undergoes a cycle in this two-level subspace, a geometric phase is accumulated. The CZ gate can be realized by choosing an appropriately time. One advantage of these parametric gates is that the gate can be realized while remaining, on average, at the DC flux bias sweet-spot. Reagor $ et$ $al.$~\cite{reagor2018demonstration} realized CZ gate via parametric control on a superconducting processor with eight qubits 
, and achieved an average process fidelity of $93\%$ for three two-qubit gates via quantum process tomography. Hong $et$ $al.$~\cite{hong2020demonstration} demonstrated a CZ operation via parametric control with average CZ fidelity as high as $98.8\%$.


\subsubsection{Resonator-induced gates }

\begin{figure}[!t]
\centering
\includegraphics[width=0.85\columnwidth]{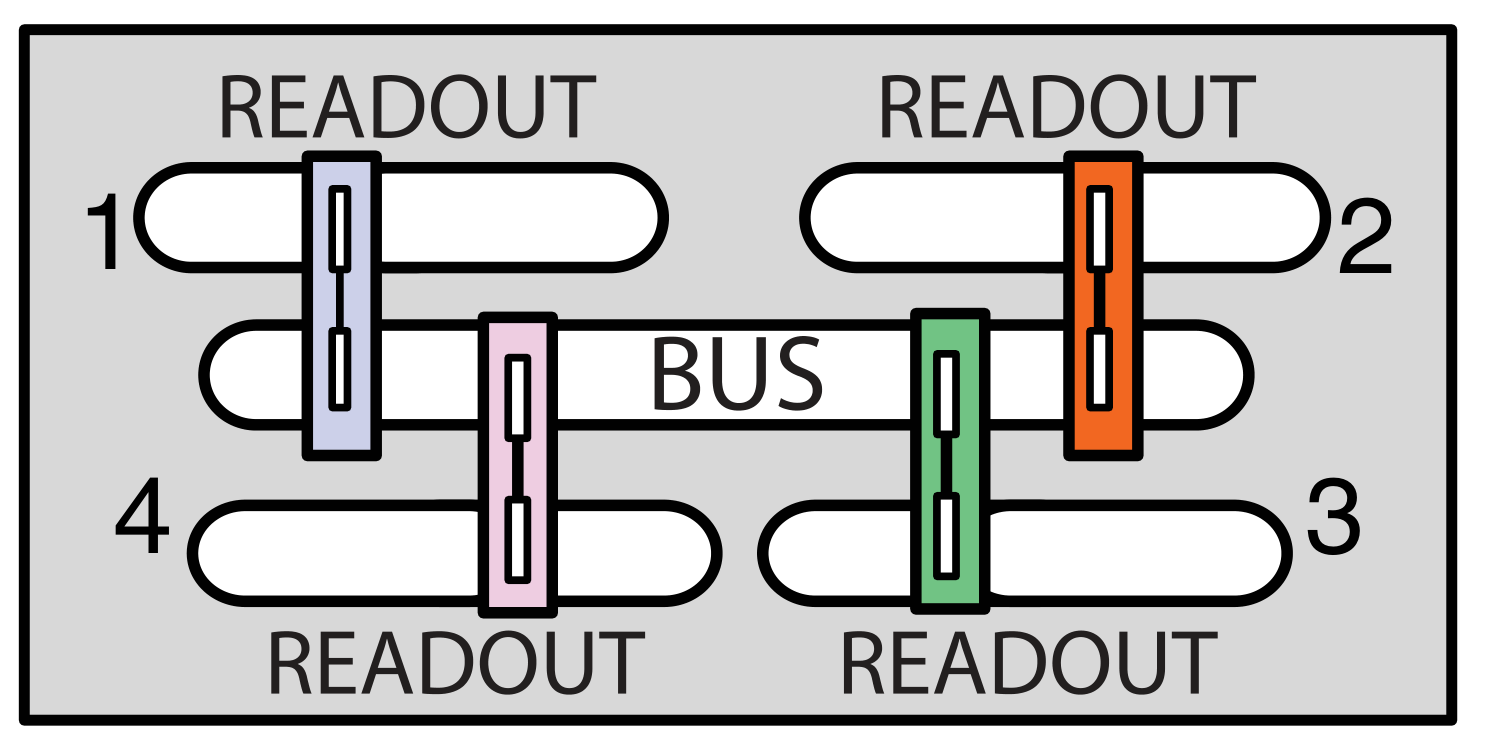}
\caption{Diagram of the 4-qubit 3D cQED system. The four qubits
are coupled with each other via common bus. Different microwave pulses input into the common cavity will cause indirect coupling of different qubit pairs. Taken from~\cite{paik2016experimental} .}
\label{fig:Figure 3-5}
\end{figure}

Resonator-induced phase gate is also an all-microwave control process multiqubit gate~\cite{paik2016experimental,puri2016high,cross2015optimized}. All qubits are statically coupled to the same driven bus resonator, which allows a high degree of flexibility in qubit frequencies. A diagram of four qubits with a resonator is shown in Fig.~\ref{fig:Figure 3-5}. The CZ gate can be realized between any pair of qubits via applying microwave to drive to the bus resonator. During the operation, the cavity state evolves from its initial vacuum state, and finally returns to vacuum state. The qubits are left unentangled from the cavity but with an acquired nontrivial phase. Once we chose an appropriate detuned by $\Delta$ to the bus, the qubit state $|1,1\rangle$ will have a $\pi$ phase relative to other states, which forms a CZ gate. In 2016, Paik $et$ $al.$~\cite{paik2016experimental} performed the CZ gate between 12 individual qubit pairs in 4-qubit superconducting 3D cQED systems using the approach of resonator-induced phase gate, and achieved fidelities from 96.55$\%$ to 98.53$\%$.

\subsubsection{Summary for two-qubit gates}

We have introduced several approaches to implement high-fidelity two-qubit gates in the superconducting quantum system. TABLE \ref{tbl:table3-1} lists several two-qubit gates implemented on superconducting quantum system in recent years.

\begin{table}[!htbp]
\centering
\caption{Two-qubit gate based on superconducting quantum system in recent years (see also the Ref.~\cite{buluta2011natural} for a similar table).}
\label{tbl:table3-1}
\begin{tabular}{|c|c|c|c|c|}
\hline
\makecell{Year\\} & \makecell{Gate type\\} & \makecell{Fidelity\\} & \makecell{Gate time\\}&\makecell{Method of \\measurement} \\
\hline
2009& CZ gate~\cite{dicarlo2009demonstration}  & 87$\%$ &NON&QST \\ 
\hline
2010& iSWAP gate~\cite{Merkel2010Generation}& 78$\%$ &NON&QST \\ 
\hline
2011& CR gate~\cite{chow2011simple}  & 81$\%$ &220ns&QPT \\ 
\hline
2012& $\sqrt{bSWAP}$ gate~\cite{poletto2012entanglement}  & 86$\%$ &800ns&QPT \\
\hline
2012& $\sqrt{iSWAP}$ gate~\cite{dewes2012characterization}  & 90$\%$ &31ns&QPT \\
\hline
2013& CZ gate~\cite{2013Microwave} & 87$\%$ &510ns&QPT \\ 
\hline
2013& CNOT gate~\cite{corcoles2013process}  & 93.47$\%$ &420ns&RB \\
\hline
2014& CZ gate~\cite{barends2014superconducting}  & 99.44$\%$ &43ns&RB \\
\hline
2014& CZ gate~\cite{chen2014qubit}  & 99.07$\%$ &30ns&RB \\
\hline
2016& CR gate~\cite{sheldon2016procedure}  & 99$\%$ &160ns&RB \\
\hline
2016& CZ gate~\cite{paik2016experimental}  & 98.53$\%$ &413ns&RB \\
\hline
2016& $\sqrt{iSWAP}$ gate~\cite{mckay2016universal}  & 98.23$\%$ &183ns&RB \\
\hline
2017& CZ gate~\cite{song2017continuous}  & 93.60$\%$ &250ns&QPT \\
\hline
2018& CZ gate~\cite{reagor2018demonstration}  & 95$\%$ &278ns&QPT \\
\hline
2018&  CZ gate~\cite{caldwell2018parametrically}  & 92$\%$ &210ns&RB \\ 
\hline
2018& iSWAP gate~\cite{caldwell2018parametrically} & 94$\%$ &150ns&RB \\ 
\hline
2018& CNOT gate~\cite{rosenblum2018cnot}  & 89$\%$ &190ns&QPT \\
\hline
2018& CNOT gate~\cite{chou2018deterministic}  & 79$\%$ &4.6$\mu$s&QPT \\
\hline
2019& CZ gate~\cite{li2019realisation}  & 99.54$\%$ &40ns&RB \\
\hline
2019& iSWAP-like gate~\cite{barends2019diabatic}  & 99.66$\%$ &18ns&XEB \\
\hline
2020& CZ gate~\cite{hong2020demonstration}  & 98.8$\%$ &176ns&RB \\
\hline
\end{tabular}
\end{table}

\subsection{Multi-qubit gates}
Multi-qubit gates for more than two qubits can also be implemented on superconducting quantum systems. Fedorov $et$ $al.$~\cite{fedorov2012implementation} implemented a Toffoli gate with three transmon qubits couple to a microwave resonator with the fidelity of 68.5$\pm$0.5$\%$ in 2012.

Song $et$ $al.$~\cite{song2017continuous} realized control-control-Z (CCZ) gate with the fidelity of 86.8$\%$ by QPT in a superconducting circuit. In 2019, Li $et$ $al.$~\cite{li2019realisation} implemented a non-adiabatic CCZ gate with the fidelity of 93.3$\%$ by using an optimised CCZ waveform. Song $et$ $al.$~\cite{song2017continuous} also realized the control-control-control-Z (CCCZ) gate with the fidelity of 81.7$\%$ by QPT.

\section{Qubit Readout}

Fast and high fidelity qubit readout is crucial in quantum computing. Historically, there are several different readout techniques for superconducting qubit, such as charge measurement~\cite{nakamura1999coherent,nakamura2002coherent}, flux measurement~\cite{mooij1999josephson,van2000quantum}, inductance measurement~\cite{vion2002manipulating}, etc.. The dispersive readout is the currently most common readout technique in the circuit QED architecture, which we will briefly introduce in this section. In addition,  we will also introduce the techniques to improve the fidelity and speed of qubit readout.

\subsection{Dispersive readout}

Dispersive readout is to obtain the qubit state information through the readout resonator~\cite{wallraff2004strong}. The qubit circuit is coupled to the readout resonator through capacitance or inductance, and the state of the qubit is detected by measuring the transmission coefficient of the readout resonator. During the measurement, the qubit and the readout resonator form the Hamiltonian of the Jaynes-Cummings model
\begin{eqnarray}
H=-\frac{\omega_q}{2}\sigma^z+\omega_ra^+a+g(\sigma^+a+\sigma^-a^+)
\label{eq4-1}
\end{eqnarray}
Considering that the absolute value of detuning $\Delta=\omega_q-\omega_r$ is much larger than the coupling strength $g$, the unitary operator
exp$[-g(\sigma^+a-\sigma^-a^+)/\Delta]$ is used to transform the Hamiltonian, and the low-level small quantities are omitted. Eq.\ref{eq4-1} becomes
\begin{eqnarray}
H=-\frac{\omega_q}{2}\sigma^z+(\omega_r+\frac{g^2}{\Delta}\sigma^z)a^+a
\label{eq4-2}
\end{eqnarray}

\begin{figure}[!t]
\centering
\includegraphics[width=0.7\columnwidth]{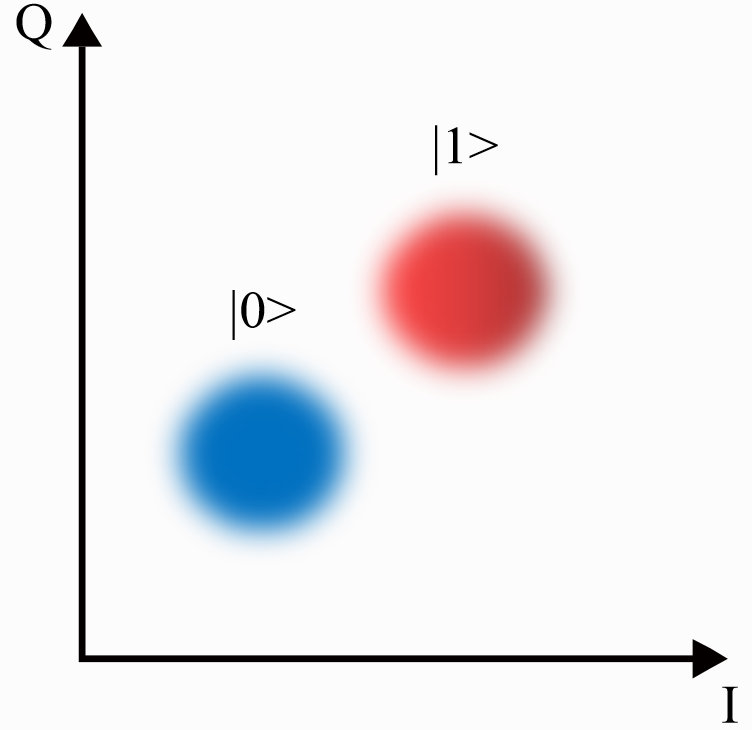}
\caption{The measurement results for $|0\rangle$ or $|1\rangle$ state.}
\label{fig:Figure 4.1}
\end{figure}

Its physical meaning is that the frequency of the readout resonator will change $\frac{g^2}{\Delta}$ when the qubit is in the $|0\rangle(|1\rangle)$
state. Let $\chi=\frac{g^2}{\Delta}$ and call $\chi$ the dispersive shift. The state of the qubit affects the frequency of the readout resonator, and the change in the frequency of the readout resonator will be reflected in the measurement of the transmission coefficient. A microwave with a specific frequency and length is input on a transmission line coupled to the readout resonator. After capturing the signal coming from the end of the transmission line and integrating the signal, we could get a point (I+iQ) in the complex plane. We can find that the measurement results of states $|0\rangle$ and $|1\rangle$ are clustered in two distinguishable clusters, respectively (see Fig.~\ref{fig:Figure 4.1}). The dispersive readout provides a feasible way for the quantum non-demolition (QND) readout of supercomducting quantum computing.

\subsection{ High-fidelity single-shot readout}

Although dispersive readout technique has been proven useful, it is often in itself insufficient to render single-shot readout performance. To achieve fast, high-fidelity single shot readout, Purcell filter and parametric amplifier are typically tools for superconducting quantum computing, which we will give a brief introduction here.

\subsubsection{Purcell filters}

In standard circuit QED, one limitation on the lifetime of the qubit is the relaxation caused by the spontaneous emission of a photon by the resonator, which is called the Purcell effect~\cite{purcell1946resonance}. Under this limitation, the lifetime of the qubit is inversely proportional to the qubit-resonator coupling strength and readout resonator bandwidth. Purcell effect is designed to suppresses spontaneous emission (at the qubit transition frequency) while being transparent at the readout resonator frequency. In recent years, various Purcell filters are proposed, and some of which are introduced below.

\begin{figure}[!t]
\centering
\includegraphics[width=0.85\columnwidth]{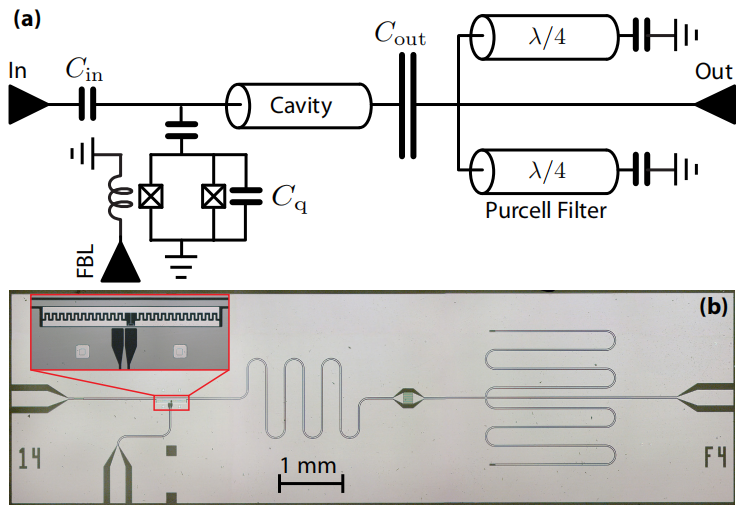}
\caption{Design and realization of the Purcell filter.  (a) Circuit model of the Purcell-filtered cavity design.  (b) Optical micrograph of the device with inset zoom on transmon qubit.Taken from~\cite{reed2010fast}.}
\label{fig:Figure 4.2}
\end{figure}

In 2010, Reed $et$ $al.$~\cite{reed2010fast} implemented the Purcell filter with a transmission-line stub terminated in an open circuit placed outside the output capacitor (see Fig.~\ref{fig:Figure 4.2}). The length of this stub was set such that it acted as a $\lambda/4$ impedance transformer to short out the 50 $\Omega$ environment at its resonance frequency. The qubit $T_1$ was improved by up to a factor of 50 compared to predicted values for an unfiltered device.

\begin{figure}[!t]
\centering
\includegraphics[width=0.85\columnwidth]{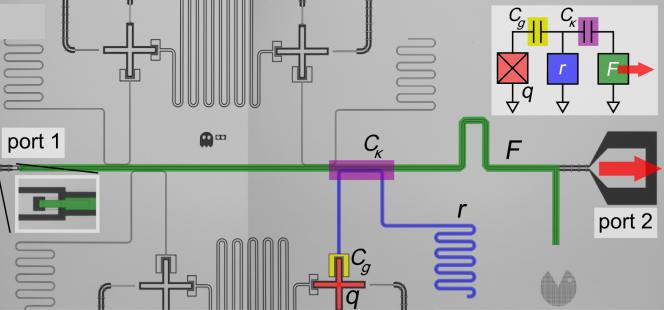}
\caption{Device layout. $C_g$ is the coupling capacitance between the qubit $q$ and the readout resonator $r$. $C_k$ is the coupling capacitance between the readout resonator and the filter resonator $F$. Taken from~\cite{jeffrey2014fast}.}
\label{fig:Figure 4.3}
\end{figure}

In 2014, Jeffrey $et$ $al.$~\cite{jeffrey2014fast} implemented bandpass Purcell filter as a quarter wave ($\lambda/4$) coplanar waveguide resonator embedded directly into the feed line (see Fig.~\ref{fig:Figure 4.3}). The bandpass filter is designed for a multiplexed measurement system that allows fast measurement without increasing environmental damping of the qubits. They demonstrated the simultaneous measurement of four qubits with intrinsic fidelities reaching 99$\%$ in less than 200 ns after the start of the measurement pulse.

\begin{figure}[!t]
\centering
\includegraphics[width=0.85\columnwidth]{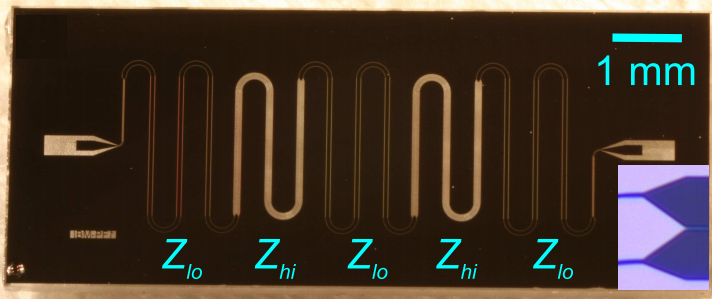}
\caption{Optical microscope image of SIPF. Inset: the transition between low-impedance and high-impedance CPW. Taken from~\cite{bronn2015broadband}.}
\label{fig:Figure 4.4}
\end{figure}

In 2015, Bronn $et$ $al.$~\cite{bronn2015broadband} implemented the stepped impedance Purcell filter (SIPF), which consists of alternating sections of high and low impedance coplanar waveguide transmission lines (see Fig.~\ref{fig:Figure 4.4}). The filter has a wide stopband and can provides protection for a large range of qubit frequencies. The qubit T1 was improved by up to a factor of 9 compared to predicted values for an unfiltered device.

In 2017, Walter $et$ $al.$~\cite{walter2017rapid} reported experimental results of 98.25$\%$ readout fidelity in only 48 ns and 99.2$\%$ readout fidelity in 88 ns, by optimizing the circuit parameters and employing a Purcell filter and phase-sensitive parametric amplification.

\subsubsection{Josephson parametric amplifier (JPA)}

Another challenge with single-shot readout is to suppress readout noise. The readout noise mainly depends on the noise temperature of the amplifier in the first stage of the readout circuit. Conventional commercial amplifier, such as high electron mobility transistors, which has a noise temperature above 2K (equivalent to the noise of tens of photons). However, the intensity of the signal often detected is only a few photons. JPA is proposed as a suitable solution for the realization of low-noise amplifiers to overcome this pressing problem.

The basic working principle of JPA is as follows. A lossless nonlinear term is generated using the Josephson effect of Josephson junctions. By applying a proper pump frequency, the energy of the pump tone is converted into the energy of the input signal through three-wave mixing or four-wave mixing, thereby outputting an amplified signal. The whole procedure is almost non-dissipative, so the noise is reduced to the level of quantum vacuum fluctuations.

The traditional JPA has relatively simple structure, and it is easy to achieve high gain. However, the bandwidth is extremely narrow, and generally does not exceed 30MHz~\cite{hatridge2011dispersive,mutus2013design}. Thus, progress in scaling to large-scale quantum computing is limited by JPA bandwidth and dynamic range. Next, we will introduce two broadband parametric amplifiers.

\subsubsection{Impedance-transformed parametric amplifier (IMPA)}

\begin{figure}[!t]
\centering
\includegraphics[width=0.85\columnwidth]{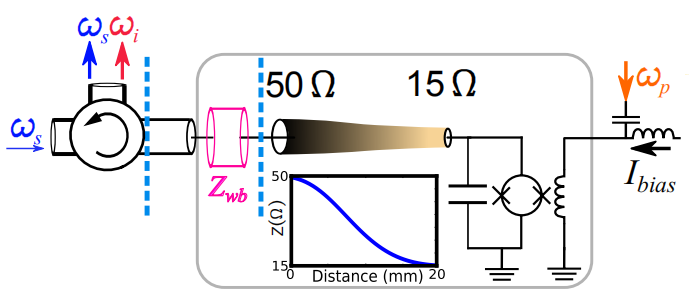}
\caption{Schematic of IMPA  (gray box).  The components to the left of the box are other microwave devices. The current bias line is used to change the frequency of IMPA and drawn on the right side of the box. Taken from~\cite{mutus2014strong}.}
\label{fig:Figure 4.5}
\end{figure}

In 2014, Mutus $et$ $al.$~\cite{mutus2014strong} prepared IMPA based on traditional JPA. The device consisted of a SQUID with a capacitor in parallel coupled to the 50 $\Omega$ environment by a tapered transmission line (see Fig.~\ref{fig:Figure 4.5}). By covering different density crossovers on different parts of the transmission line, the characteristic impedance of the transmission line smoothly changed from 50 $\Omega$ to 15 $\Omega$. By changing the characteristic impedance of the transmission line, the parametric amplifier achieved a 700 MHz amplification bandwidth, a gain of more than 15 dB, a saturation power of -103 dBm, and a noise temperature of 200 mK.

\subsubsection{Traveling wave parametric amplifier (TWPA)}

\begin{figure}[!t]
\centering
\includegraphics[width=0.85\columnwidth]{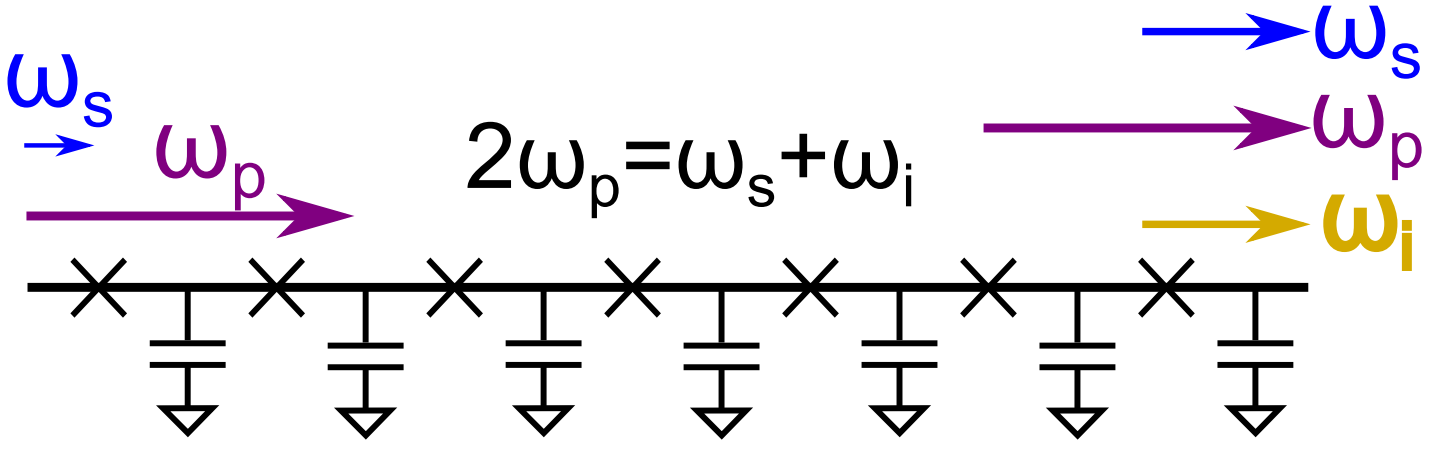}
\caption{Schematic of TWPA. Thousands of Josephson junctions are connected in series to amplify the input signal. Taken from~\cite{white2015traveling}.}
\label{fig:Figure 4.6}
\end{figure}

In 2015, White $et$ $al.$~\cite{white2015traveling} prepared TWPA, consisting of thousands of Josephson junctions in series, which formed a 50 $\Omega$ lumped element transmission line by connecting capacitors in parallel (see Fig.~\ref{fig:Figure 4.6}). Continuous phase matching can be achieved by inserting $\lambda/4$ resonators at fixed intervals on the transmission line to correct the pump phase, which can increase the gain exponentially with minimal manufacturing complexity. In this case, the gain can be increased by simply increasing the length of the device. The TWPA achieved a 4 GHz amplification bandwidth, an average gain of 12 dB, a saturation power of -99 dBm, and a noise temperature of 600 mK.

\subsection{Real-time quantum feedback}

\begin{figure}[!t]
\centering
\includegraphics[width=0.85\columnwidth]{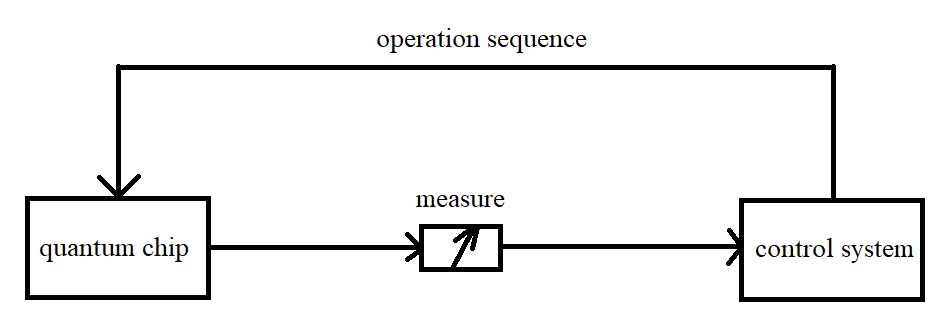}
\caption{Schematic of quantum feedback control.}
\label{fig:Figure 4.7}
\end{figure}

Real-time quantum feedback (see Fig.~\ref{fig:Figure 4.7}) is highly sought-after due to its potential for quantum error correction (QEC) and maintaining quantum coherence~\cite{zhang2017quantum}. The realization of quantum feedback requires that the entire quantum chip measurement and control system must have excellent overall coordination and extremely fast response time. The whole procedure, including readout operations, analysis of readout data, and generation of feedback operations, must be completed before the decoherence of qubits. Analog feedback schemes such as those reported in Refs.~\cite{vijay2012stabilizing,campagne2016using} feature feedback latencies on the order of 100 ns, where the latencies are limited by analog bandwidth and delays in the cables in the cryogenic setups. However, analog signal processing circuits have limited flexibility. In 2018, Salath{\'e} $et$ $al.$~\cite{salathe2018low} implemented an FPGA-based system which offers versatile and convenient programming. The feedback-capable signal analyzer allows for real-time digital demodulation of a dispersive readout signal and the generation of a qubit-state-dependent trigger with input-to-output latency of 110 ns. However, they only implemented quantum feedback control of a single qubit. Further research on the quantum feedback control of multiple qubits is highly needed in the future.

\section{Quantum Error Correction}

Although tremendous progress on experimental techniques has been achieved with superconducting quantum computing system in recent years, including coherence time, gate fidelity, and readout fidelity, quantum error correction is still necessary for large-scale quantum computations. In this section, we will introduce some typical quantum error correction schemes.

\subsection{surface code}

\begin{figure}[!t]
\centering
\includegraphics[width=0.8\columnwidth]{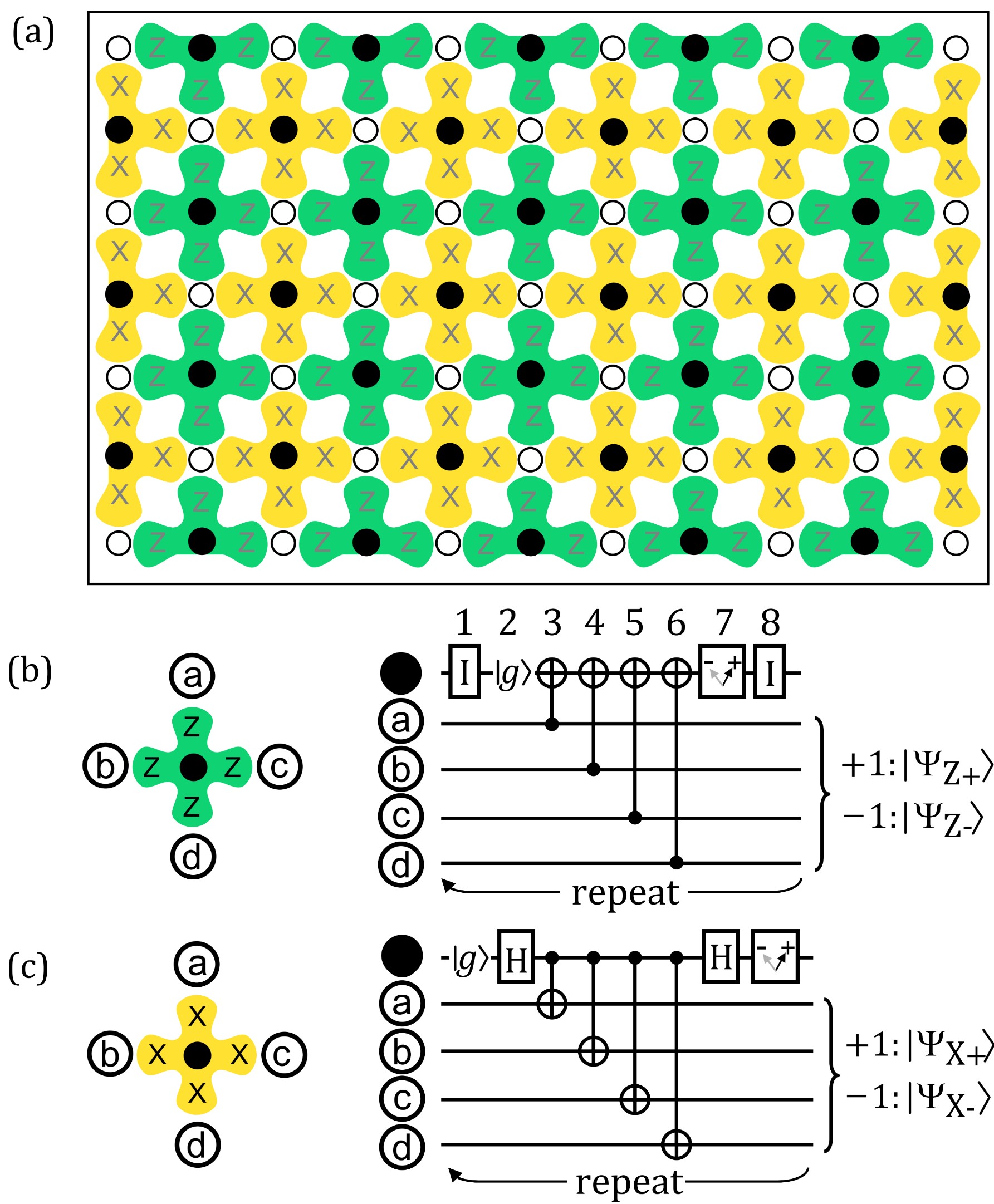}
\caption{(a) 2D surface code. Data qubits and measurement qubits are open circles and filled circles, respectively. (b) Geometry and quantum circuit for a measure-$Z$ qubit. (c) Geometry and quantum circuit for a measure-$X$ qubit. Taken from~\cite{fowler2012surface}.}
\label{fig:Figure.5-1}
\end{figure}

Surface code represent a promising route towards universal and scalable fault-tolerant quantum computation, due to its nearest-neighbour qubit layout and high error rate thresholds ~\cite{fowler2012surface}. As shown in Fig.~\ref{fig:Figure.5-1}, each data qubit (represented by open circles) contacts four measurement qubits (represented by filled circles), and each measurement qubit also contacts four data qubits. There are two types of measurement qubits, ``measure-$Z$" qubits and ``measure-$X$" qubits, which are colored by green and orange in Fig.~\ref{fig:Figure.5-1}(a), respectively. The central task during the error correction is to perform 4-qubit parity measurements using $ZZZZ$ and $XXXX$ stabilizer operators to check whether there has been a bit-flip or a phase-flip in the data qubit.

The repetition code is an one-dimensional variant of the surface code, which is able to protect against bit-flip errors or phase-flip errors, but cannot simultaneously detect both bit- and phase-flip errors. A three-qubit repetition code was first implemented by Reed $et$ $al.$ ~\cite{reed2012realization} in a 3 transmon qubit circuit without ancillas, correcting a single error using a Toffoli gate. Riste $et$ $al.$~\cite{riste2015detecting} realized a three-qubit repetition code using 5 qubits to detect bit-flip errors in a logical qubit, where 3 qubits are used for encoding, and 2 ancilla qubits are used for the syndrome. Chow $et$ $al.$ demonstrated a high-fidelity parity detection of two code qubits via measurement of a third syndrome qubit by using the `cross-resonance' two-qubit gate~\cite{chow2014implementing}. In 2014, Barends $et$ $al.$~\cite{barends2014superconducting} demonstrated a two-qubit gate fidelity of up to 99.4$\%$ using Xmon qubits, which is the first time that the fidelity of two-qubit gate surpassed the error threshold for the surface code. Moreover, using the similar gate control technique, Kelly $et$ $al.$~\cite{kelly2015state} extended their system to a linear 1D chain with 9 qubits, and implemented both a five- and nine-qubit repetition code, providing a promising step toward the 2D-surface code.

2D-surface code could simultaneously detect both bit- and phase-flip errors. In 2015, C\'{o}rcoles $et$ $al.$ first demonstrated a quantum error detection protocol on a two-by-two planar lattice of superconducting qubits~\cite{corcoles2015demonstration}. The 4-qubit parity measurements, a basic operation in 2D-surface code scheme, were demonstrated by Takita $et$ $al.$ ~\cite{takita2016demonstration} . In order to promote the practicality of 2D-surface code, more advanced experimental techniques are needed, such as real-time feedback, repeated detection, and so on. Andersen $et$ $al.$ demonstrated the real-time stabilization of a Bell state with a fidelity of $F\approx74\%$ in up to 12 cycles of the feedback loop ~\cite{kraglund2019entanglement} , and implemented a seven-qubit surface code for repeated quantum error detection~\cite{andersen2019repeated}.

\begin{figure}[!t]
\centering
\includegraphics[width=0.85\columnwidth]{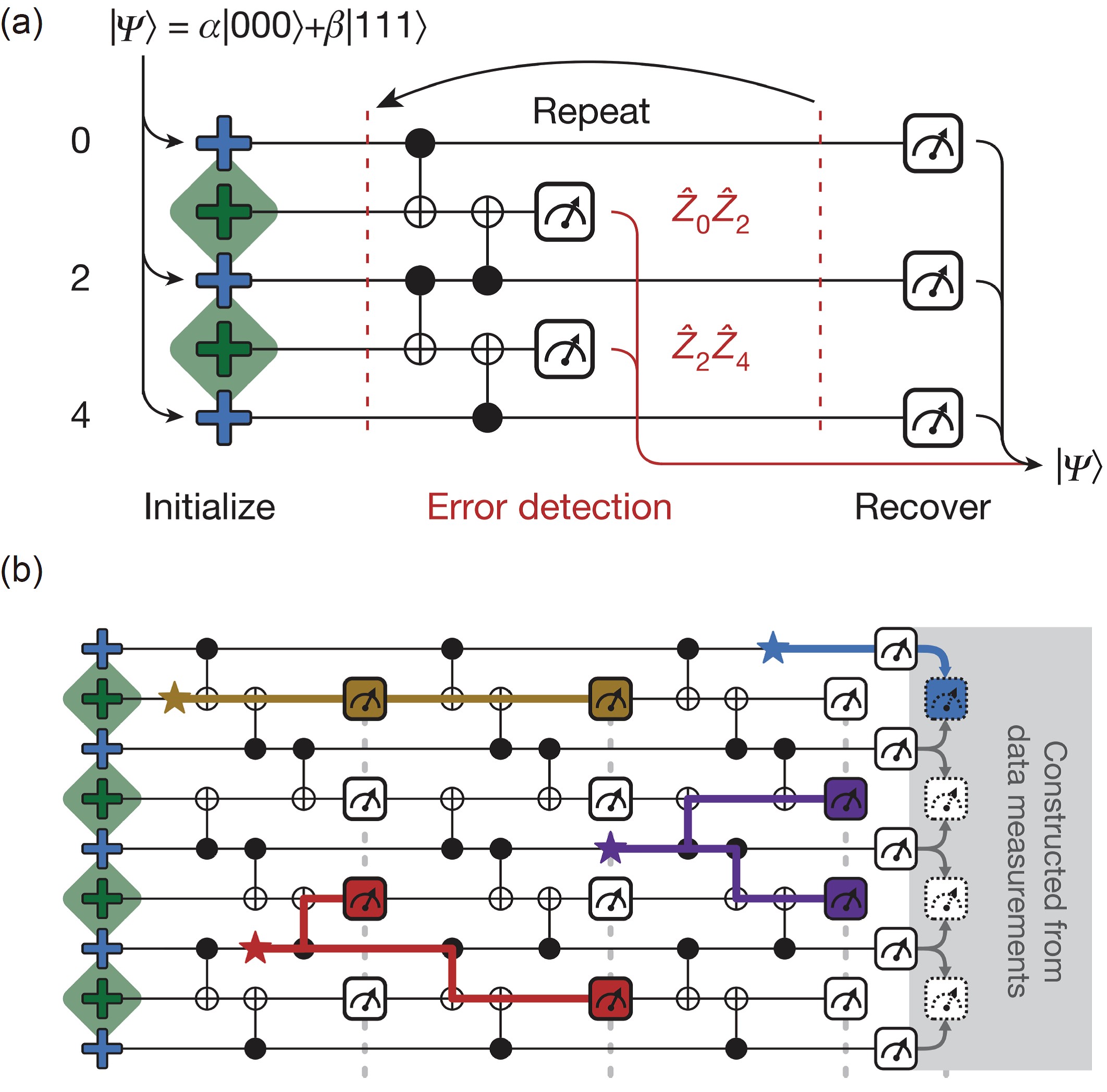}
\caption{(a) The five qubits repetition code. (b) The quantum circuit for three cycles of the nine-qubit repetition code. Taken from~\cite{kelly2015state}.}
\label{fig:Figure.5-2}
\end{figure}

\subsection{Bosonic codes}

Bosonic codes have recently arisen as a hardware-efficient route to implementing quantum error correction by taking advantage of the infinite dimensionality of a bosonic Hilbert space, rather than to duplicate many two-level qubits as in the surface code. Until now, a broad class of bosonic error-correcting codes has been proposed. For example, the well-known cat codes ~\cite{cochrane1999macroscopically,leghtas2013hardware} and binomial codes ~\cite{michael2016new}, which are inspired by the Gottesman, Kitaev, and Preskill (GKP)~\cite{gottesman2001encoding} proposal that encoding a qubit in an oscillator. Both GKP codes and cat codes are formed using superpositions of an infinite number of Fock states, while the binomial code states are formed from a finite superposition of Fock states weighted with square roots of binomial coefficients.

\begin{figure}[!t]
\centering
\includegraphics[width=0.9\columnwidth]{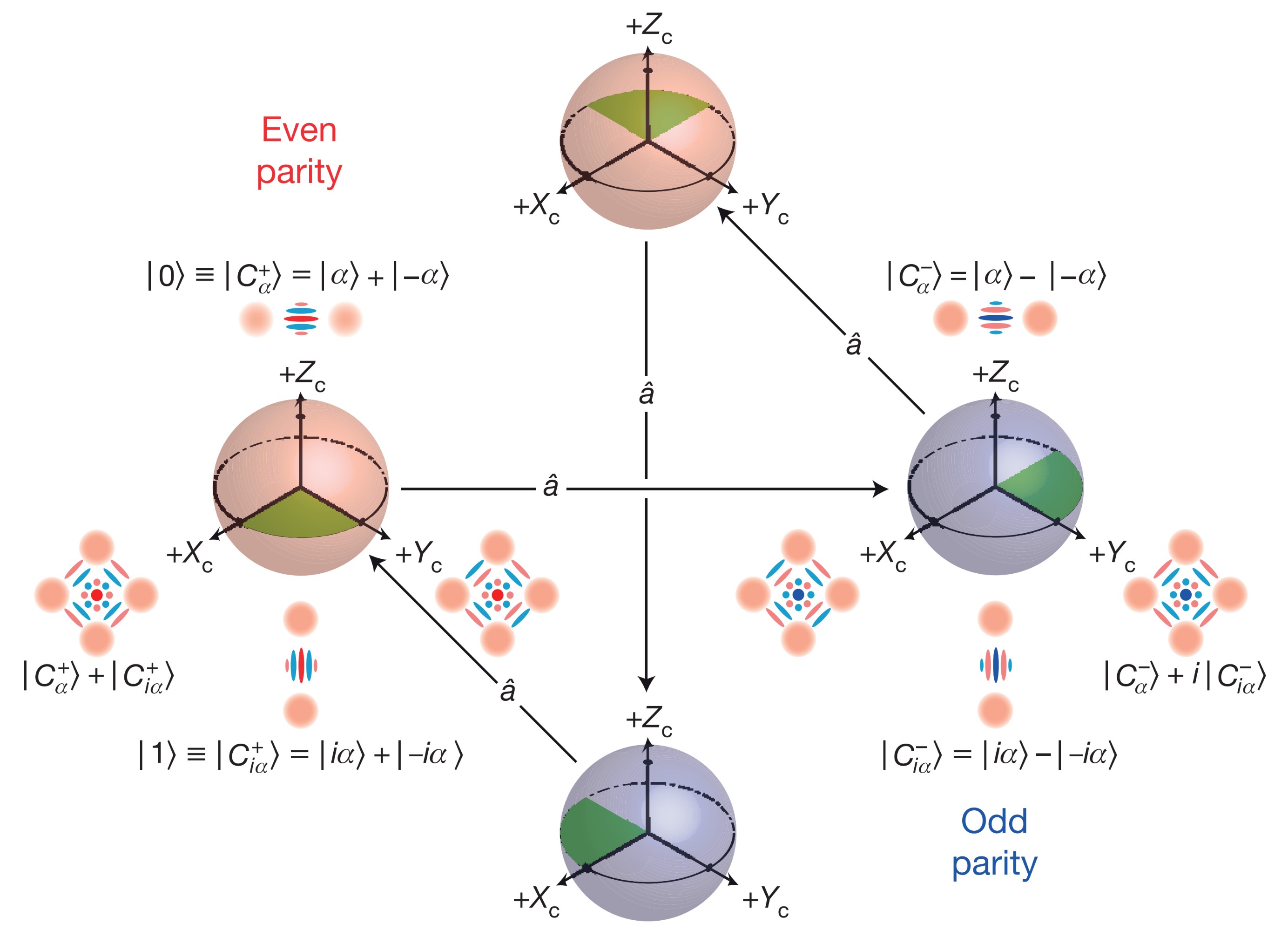}
\caption{The cat-code cycle. Taken from~\cite{ofek2016extending}.}
\label{fig:Figure.5-3}
\end{figure}

Recent experimental results for the bosonic codes show great promise. In 2015, Leghtas $et$ $al.$ shown that engineering the interaction between a quantum system and its environment can induce stability for the delicate quantum states~\cite{leghtas2015confining} , which provides the possibility toward robustly encoding quantum information in multidimensional steady-state manifolds. Wang $et$ $al.$ demonstrated a Schr${\rm{\ddot o}}$dinger's cat that lives in two cavities ~\cite{wang2016schrodinger} . The break-even point of QEC is when the lifetime of a qubit exceeds the lifetime of the constituents of the system. Ofek $et$ $al.$ first demonstrated a QEC system that reaches the break-even point by suppressing the natural errors using the bosonic cat code~\cite{ofek2016extending} (see Fig.~\ref{fig:Figure.5-3}). Hu $et$ $al.$  experimentally demonstrated repetitive QEC approaching the break-even point of a single logical qubit encoded in a hybrid system consisting of a superconducting circuit and a bosonic cavity using a binomial bosonic code~\cite{hu2019quantum} . A critical component of any quantum error-correcting scheme is detection of errors by using an ancilla system, Rosenblum $et$ $al.$ demonstrated a fault-tolerant error-detection scheme that suppresses spreading of ancilla errors by a factor of 5, while maintaining the assignment fidelity~\cite{rosenblum2018fault} .

Showing the feasibility of operations on logical qubits is a necessary step towards universal QEC. Heeres $et$ $al.$ demonstrated a high-fidelity implementation of a universal set of gates on a qubit encoded into an oscillator using the cat-code~\cite{heeres2017implementing}. Recently, Reinhold $et$ $al.$ presented an ancilla-enabled gate that uses the principle of path-independence to maintain coherence of the logical qubit in the presence of ancilla errors, and demonstrated that the error-correction suppresses the propagation of the dominant errors~\cite{reinhold2019error}. Beyond single logical qubit operation, a high-fidelity entangling gate between multiphoton states encoded in two cavities~\cite{rosenblum2018cnot} and the exponential-SWAP unitary ~\cite{gao2018entangling} have been realized.

\subsection{Other error-correction method}

\begin{figure}[!t]
\centering
\includegraphics[width=1\columnwidth]{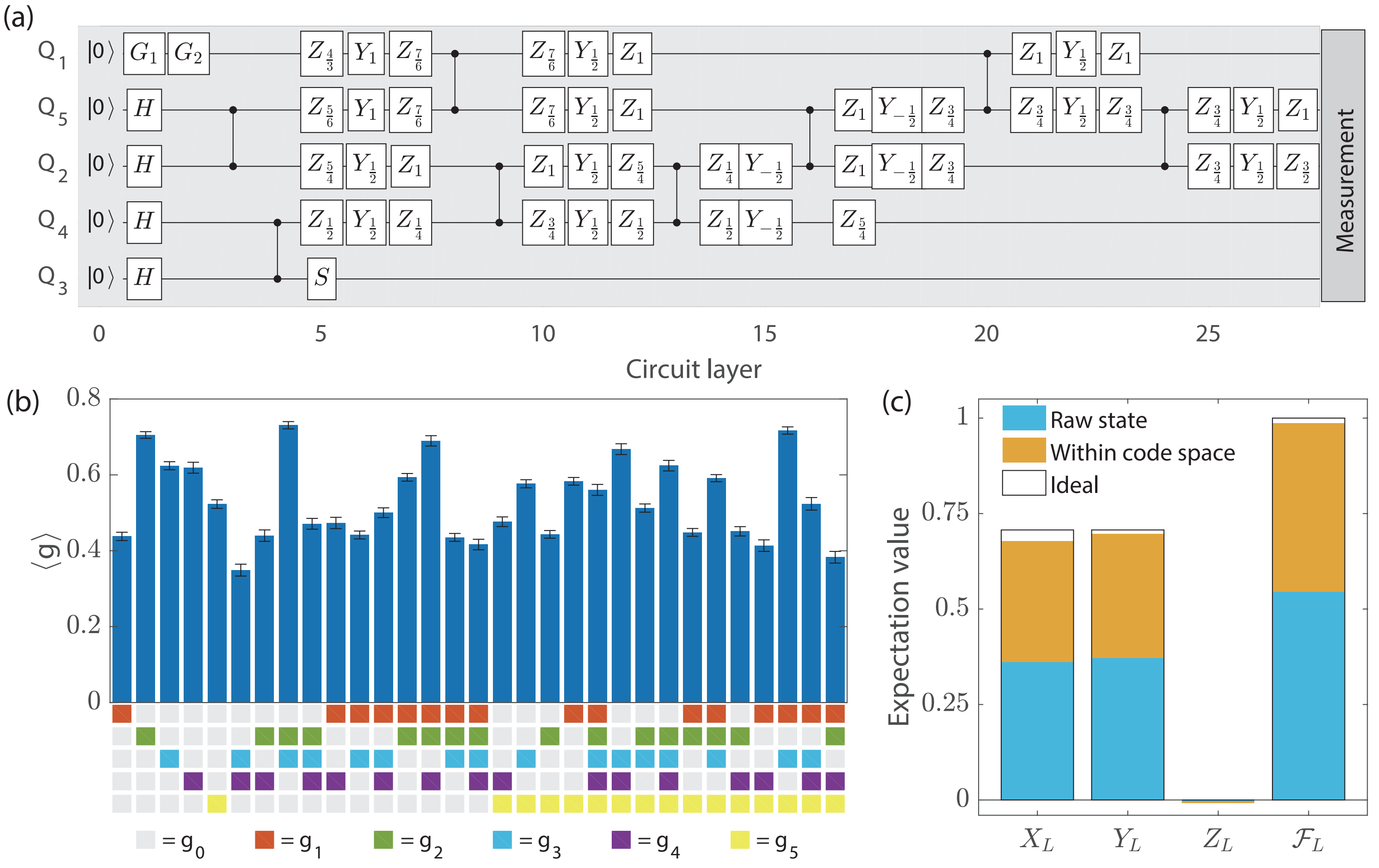}
\caption{(a) Quantum circuit for the five-qubit code. (b) Expectation values of 31 stabilizers for the encoded logical state ${\rm{|}}T{\rangle _L} = {(|0\rangle _L} + {e^{i\pi /4}}|1{\rangle _L})/\sqrt 2$. (c) Expectation values of logical Pauli operators and state fidelity of the encoded magic state ${\rm{|}}T{\rangle _L}$. Taken from~\cite{gong2019experimental}.}
\label{fig:Figure.5-4}
\end{figure}

Besides surface code and bosonic codes, some other quantum error correction codes have also been experimentally demonstrated on superconducting system. Takita $et$ $al.$ implemented the [4,2,2] code in a five-qubit superconducting transmon device and characterize output states produced by fault-tolerant state preparation circuits ~\cite{takita2017experimental}. Harper $et$ $al.$ observed an order of magnitude improvement in the infidelity of the gates, with the two-qubit infidelity dropping from 5.8(2)$\%$ to 0.60(3)$\%$, by running a randomized benchmarking protocol in the logical code space of the [4,2,2] code~\cite{harper2019fault}. Moreover, Ming $et$ $al.$ emploiyed an array of superconducting qubits to realise the [5,1,3] code for several typical logical states including the magic state, an indispensable resource for realising non-Clifford gates (see Fig.~\ref{fig:Figure.5-4}) ~\cite{gong2019experimental} . All of these experiments have shown the importance of quantum error correction. However, in order to achieve fault-tolerant quantum computing in the future, we still need to make more efforts in experimental techniques.

\section{Algorithm Implementation}

Currently, quantum computer devices are still small scale, and their capabilities have not reached the level beyond small demonstration algorithms. However, small scale demonstrations are crucial for the realization of more advanced experimental techniques, since it can provide useful insights into the performance of existing systems and the role of architecture in quantum computer design. Superconducting quantum computer is fully programmable multi-qubit machines that could in principle provide the user with the flexibility to implement arbitrary quantum circuits. Thus, a number of important quantum algorithms have been demonstrated using superconducting quantum system, especially after IBM made several quantum devices available to the public through its quantum cloud service. In this section, we are not going to introduce all of these works. Instead, here we only provide some examples to figure out what superconducting quantum computers will be able to do when we build them.

\subsection{Quantum simulation}

``Nature isn't classical, dammit, and if you want to make a simulation of nature, you'd better make it quantum mechanical,'' the physicist Richard Feynman famously quipped in 1981. ``And by golly it's a wonderful problem, because it doesn't look so easy.'' Although simulating a complex quantum system is known to be a difficult computational problem, this difficulty may be overcome by using some well-controllable quantum system to study another less controllable or accessible quantum system. In the past decades, the tremendous progress in controllable quantum systems has made the physical implementation of quantum simulation a reality, and we are expected to apply quantum simulation to study many issues, such as condensed-matter physics, high-energy physics, atomic physics, quantum chemistry and cosmology \cite{buluta2009quantum,georgescu2014quantum}. Compared with universal quantum computing, the quantum simulator would be easier to construct since less control are required.

\begin{figure}[!t]
\centering
\includegraphics[width=1\columnwidth]{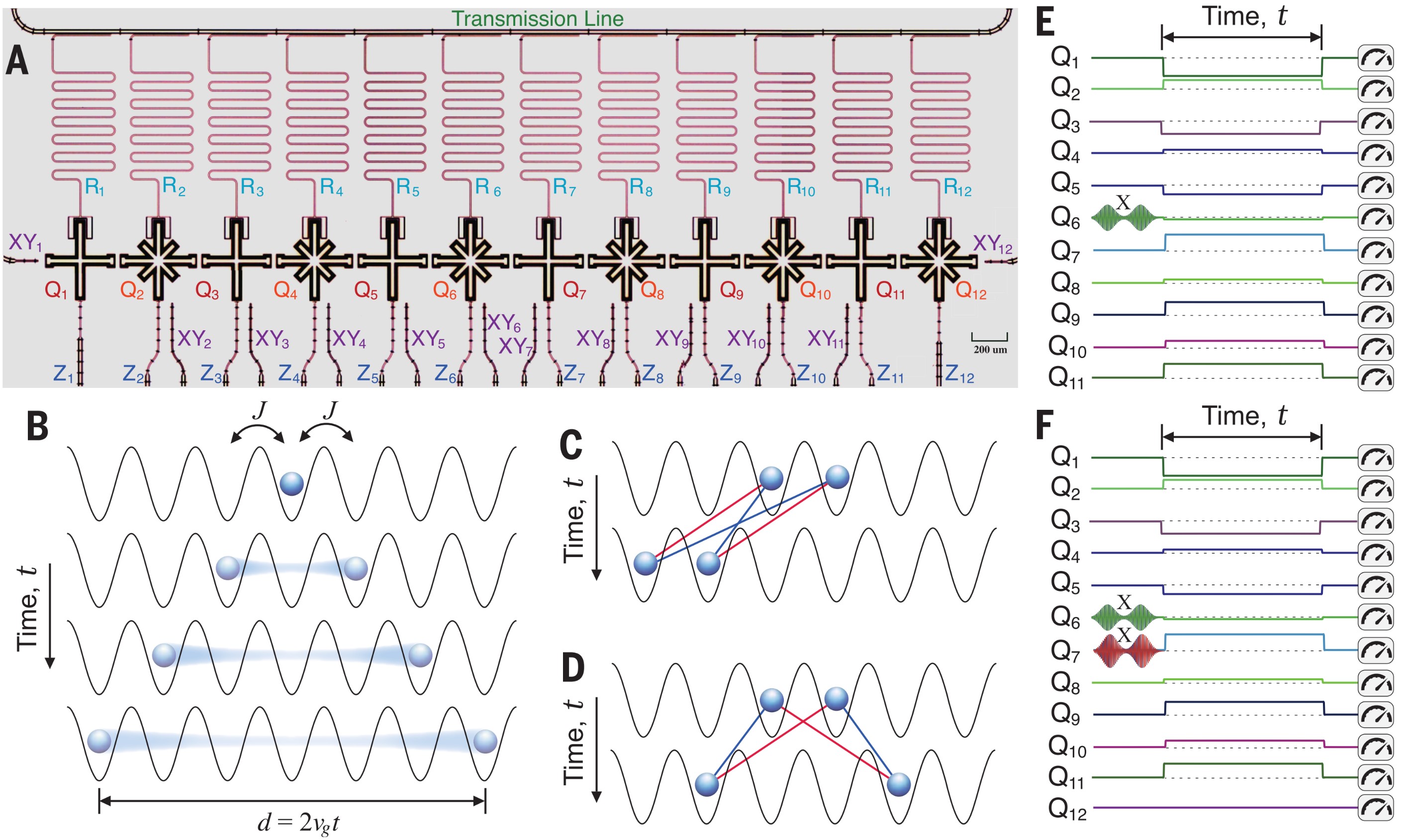}
\caption{Quantum walks of one and two photons in a 1D lattice of a superconducting processor. (a) Optical micrograph of the 12-qubit chain. (b) One photon quantum walk. (c) and (d) are quantum walk for two weakly interacting photons and two strongly interacting photons, respectively. (e) and (f) are the experimental waveform sequences for single-photon quantum walk and two-photon quantum walk, respectively. Taken from~\cite{yan2019strongly}.}
\label{fig:Figure 6.1}
\end{figure}

\begin{figure}[!t]
\centering
\includegraphics[width=1\columnwidth]{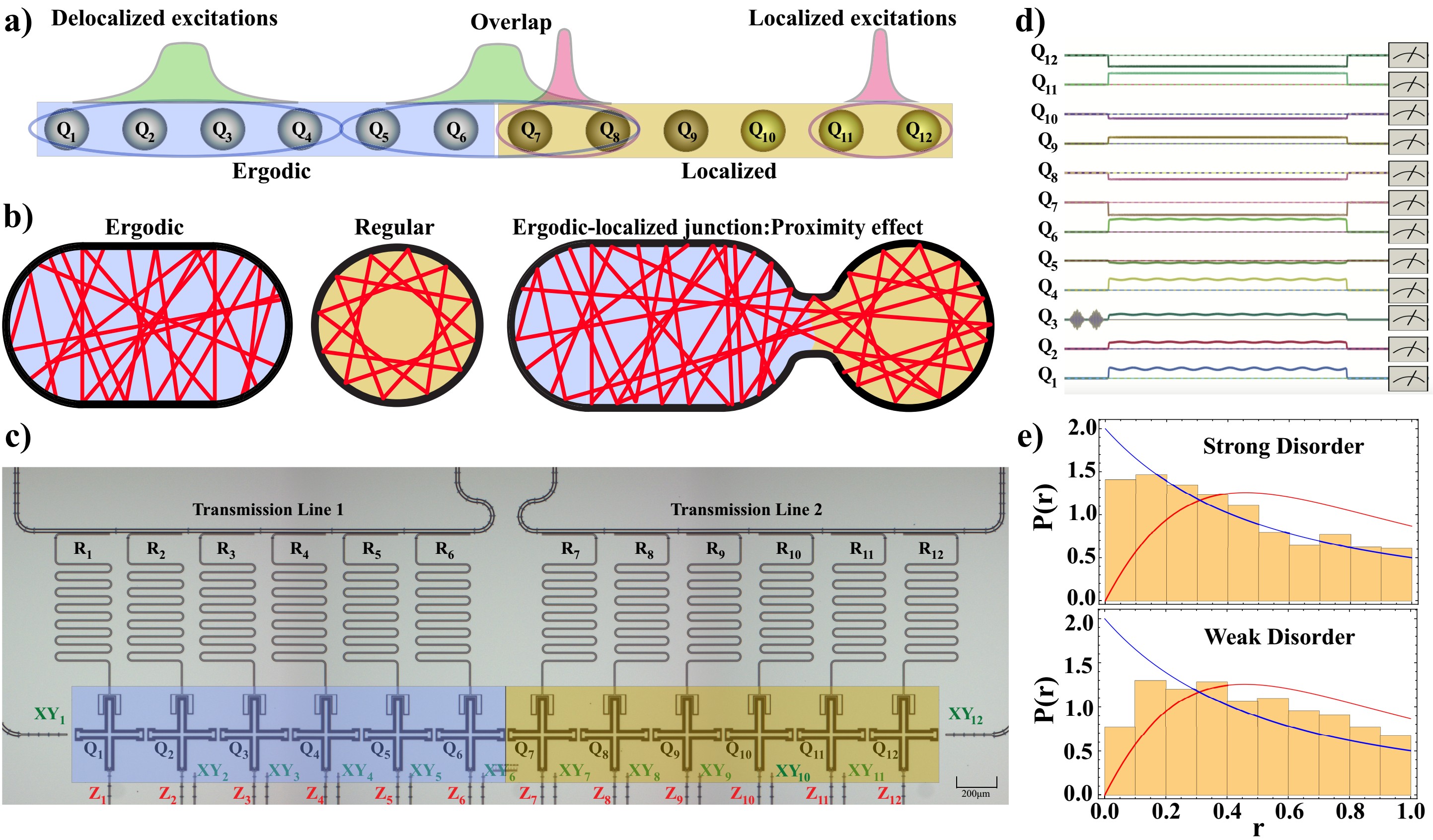}
\caption{Ergodic-localized junction with superconducting qubits. (a) when disordered and driven domains are coupled, localization can be destroyed due to the overlap between localized and delocalized states. (b) Depicts stadium and circular billiards, which exhibit ergodic and regular behavior, respectively. (c) Optical micrograph of the superconducting chip. (d) Experimental waveform sequences to generate the ergodic-localized junctions. (e) The quasienergy level statistics of the ergodic-localized junction for an array of $L=12$ qubits. Taken from~\cite{zha2020ergodic}.}
\label{fig:Figure 6.2}
\end{figure}

In order to emulate the time evolution of the quantum system, the Hamiltonian of the quantum simulator should to be very similar to the quantum system to be simulated. The superconducting transmon qubits of the Xmon variety can be described by the Bose-Hubbard model, which makes simulating the Hamiltonian related to the Bose-Hubbard model become possible. Roushan $et$ $al.$ implemented a technique for resolving the entire energy spectrum of a Hamiltonian on a linear array of Transmon qubits~\cite{roushan2017spectroscopic}. And recently, Wozniakowski $et$ $al.$ introduced a approach for predicting the measured energy spectrum~\cite{wozniakowski2020boosting}, which may be used for quantum computer calibration. Yan $et$ $al.$ experimentally implemented quantum walks of one and two strongly interacting photons in a one-dimensional array of superconducting qubits with short-range interactions, and observed the light cone-like propagation of quantum information, especially entanglement, and the photon antibunching with the two-photon HBT interference ~\cite{yan2019strongly} (see Fig.~\ref{fig:Figure 6.1}). A Bose-Hubbard ladder with a ladder array of 20 qubits on a 24-qubit superconducting processor was constructed by Ye $et$ $al.$, and the quench dynamics of single- and double-excitation states with distinct behaviors were studied on their system ~\cite{ye2019propagation}. Zha $et$ $al.$ performed an ensemble average over 50 realizations of disorder to clearly shows the proximity effect by using a superconducting quantum processor ~\cite{zha2020ergodic} (see Fig.~\ref{fig:Figure 6.2}). Other models, such as spin model ~\cite{xu2018emulating}, quantum Rabi model ~\cite{forn2010observation,yoshihara2017superconducting,braumuller2017analog,romero2012ultrafast,kockum2019ultrastrong}, and Lipkin-Meshkov-Glick model~\cite{xu2020probing} were also investigated.

\begin{figure}[!t]
\centering
\includegraphics[width=\columnwidth]{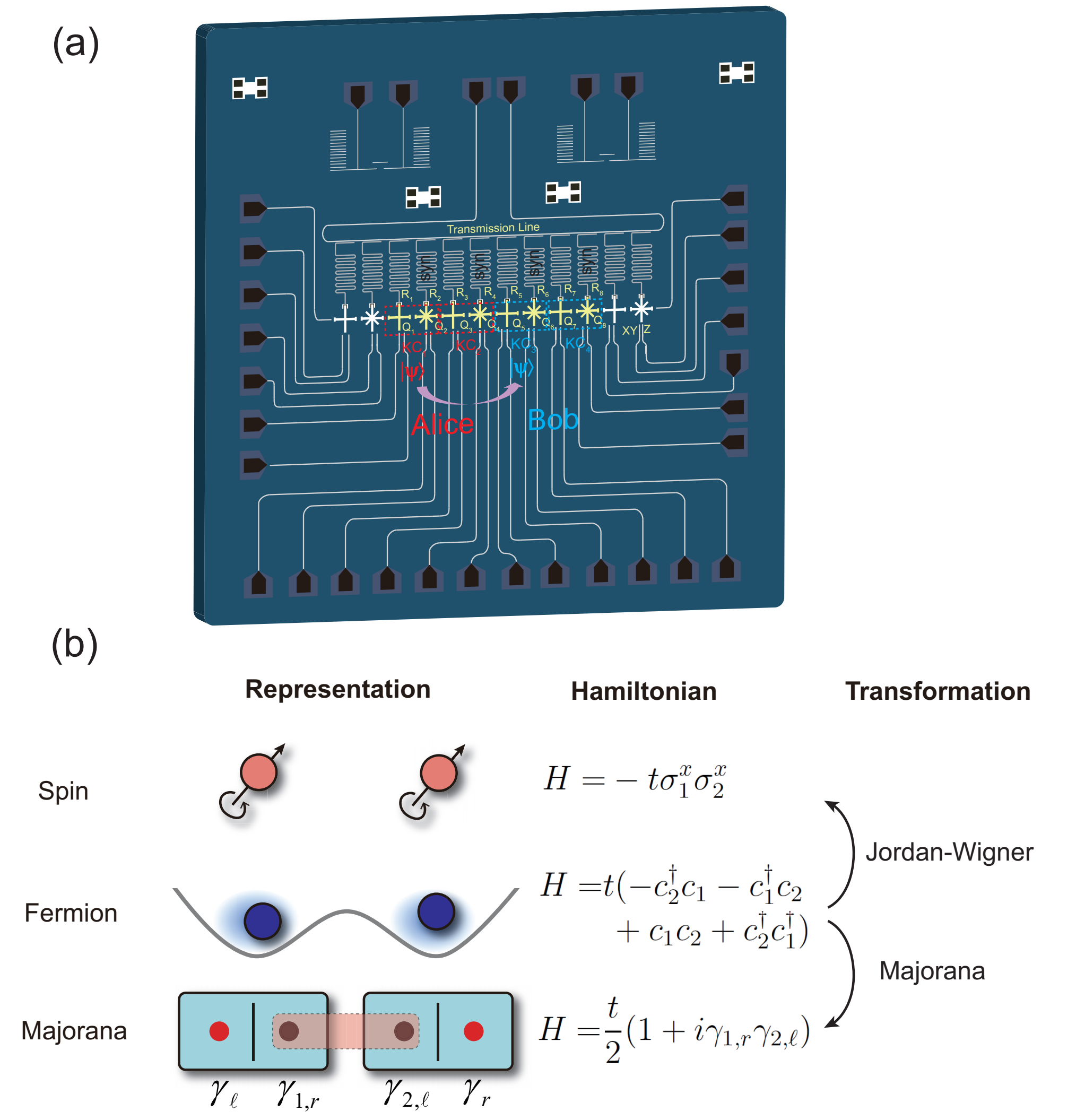}
\caption{Experimental configuration and encoding of quantum states in a Kitaev chain. (a) The superconducting quantum processor. Qubits $Q_1$ to $Q_4$ are held by Alice, qubits $Q_5$ to $Q_8$ are held by Bob. Pairs of qubits form a Kitaev chain ($KC$), each which encode a single logical qubit. An encoded qubit is teleported from $KC_1$ to $KC_3$. (b) Mapping between spin, fermions, and Majorana modes. Taken from~\cite{huang2020quantum}.}
\label{fig:MZM}
\end{figure}

\begin{figure}[!t]
\centering
\includegraphics[width=0.75\columnwidth]{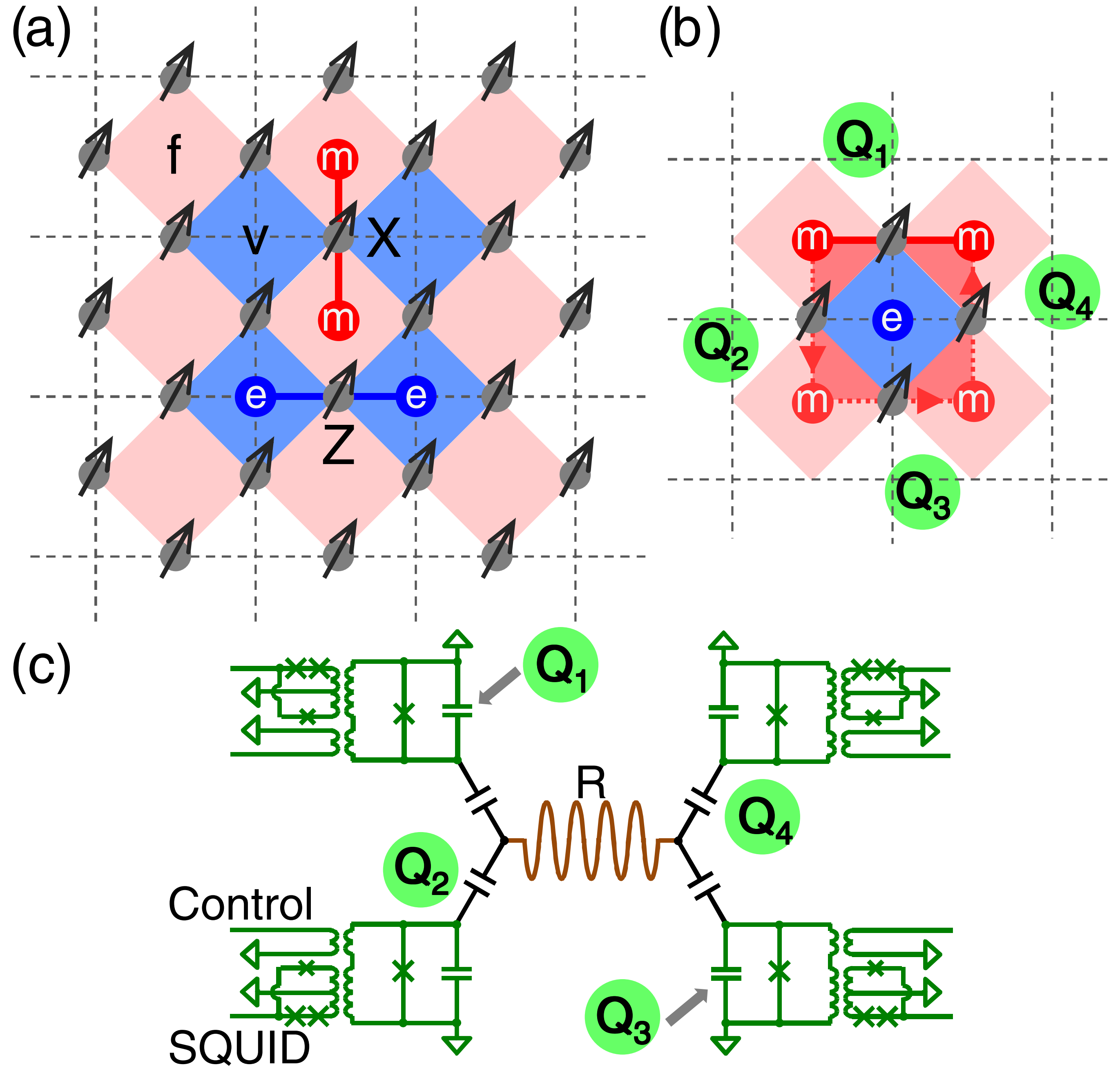}
\caption{(a) Illustration of the toric code model. (b) The minimal unit of the toric code using for qubits. (c) Schematic of the superconducting circuit featuring four qubits coupled to a central resonator. Taken from~\cite{zhong2016emulating}.}
\label{fig:Figure 6.3}
\end{figure}

Although the analog quantum simulation that uses a well-controllable quantum system to simulate the time evolution of another complex quantum system with a similar Hamiltonian seems very natural, the types of systems that can be simulated are limited. Digital quantum simulation is not constrained by the types of system, which is an approach for simulating the evolution of a quantum system based on gate model. This approach relies on decomposing the unitary operations that describe the time evolution of Hamiltonians into a set of quantum gates that can be implemented on the simulator hardware, and Trotter decomposition is one of the most popular methods to accomplish this task ~\cite{abrams1997simulation,aspuru2005simulated,whitfield2011simulation}. Different types of quantum systems, such as quantum Rabi model ~\cite{langford2017experimentally}, spin model ~\cite{salathe2015digital,las2014digital,barends2016digitized} and fermionic model ~\cite{barends2015digital,o2016scalable,las2015fermionic}, were investigated with the digital quantum simulation by different research groups. Moreover, this gate-based approach could also been used for simulating the behavior of anyons, which are exotic quasiparticles obeying fractional statistics ranging continuously between the Fermi-Dirac and Bose-Einstein statistics ~\cite{wilczek1982quantum,you2010quantum,han2007scheme}. Zhong $et$ $al.$ presented an experimental emulation of creating anionic excitations in a superconducting circuit that consists of four qubits ~\cite{zhong2016emulating} (see Fig.~\ref{fig:Figure 6.3}). Song $et$ $al.$ presented an experiment of demonstrating the path independent nature of anyonic braiding statistics with a superconducting quantum circuit ~\cite{song2018demonstration}, and a similar work was carried out with linear optical system ~\cite{liu2019demonstration}. In 2020, Huang $et$ $al.$ demonstrated in a superconducting quantum processor that the spin-mapped version of the Majorana zero modes can be used to perform quantum teleportation ~\cite{huang2020quantum} (see Fig.~\ref{fig:MZM}). Besides, Refs.~\cite{kockum2018decoherence, kannan2019waveguide} show an architecture opening up new possibilities for quantum simulation and computation with superconducting qubits.

\subsection{Quantum algorithm demonstration}

\begin{figure}[!t]
\centering
\includegraphics[width=1\columnwidth]{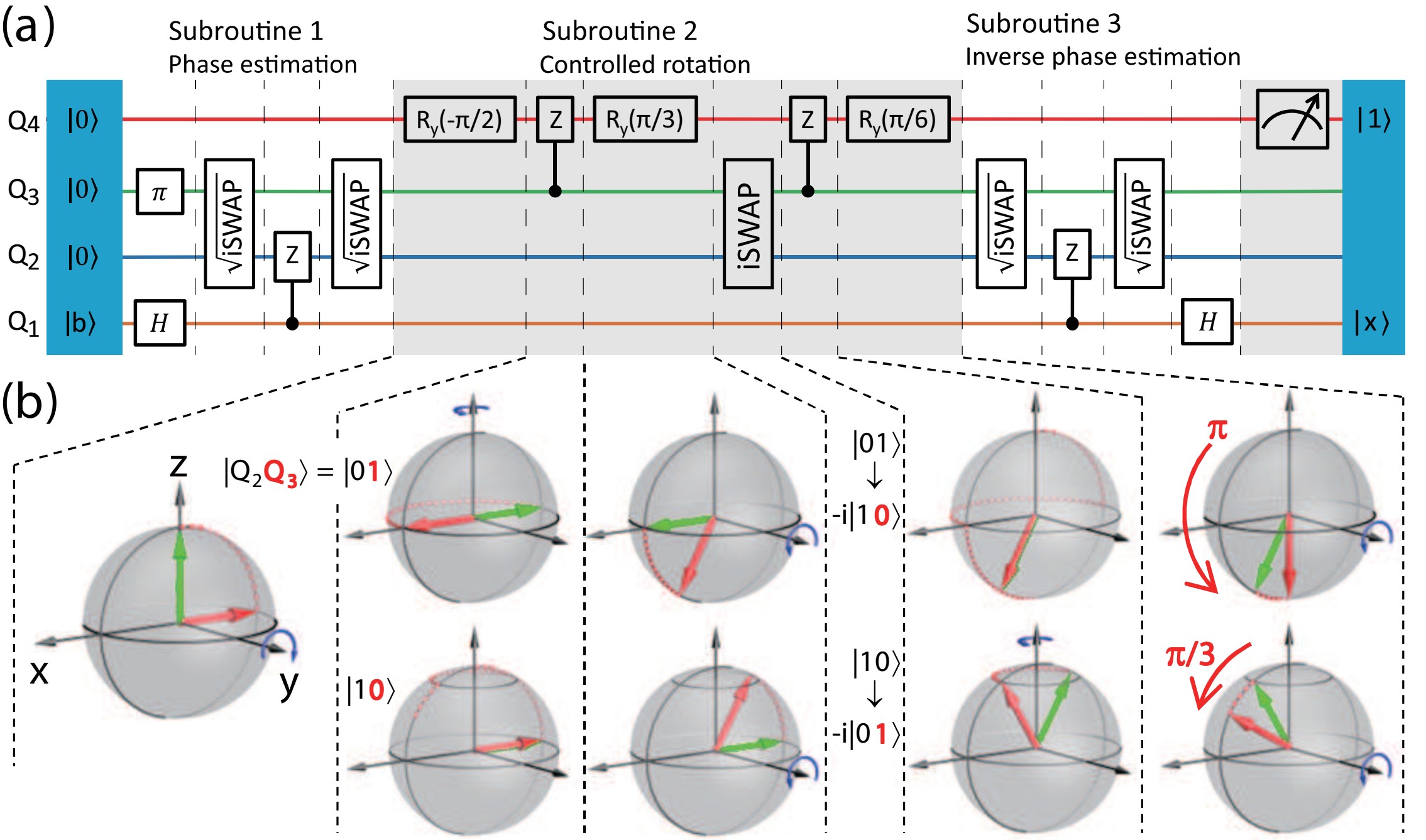}
\caption{Compiled quantum circuits for solving $2 \times 2$ linear equations using four qubits.Taken from~\cite{zheng2017solving}.}
\label{fig:Figure 6.4}
\end{figure}

\begin{figure}[!t]
\centering
\includegraphics[width=0.85\columnwidth]{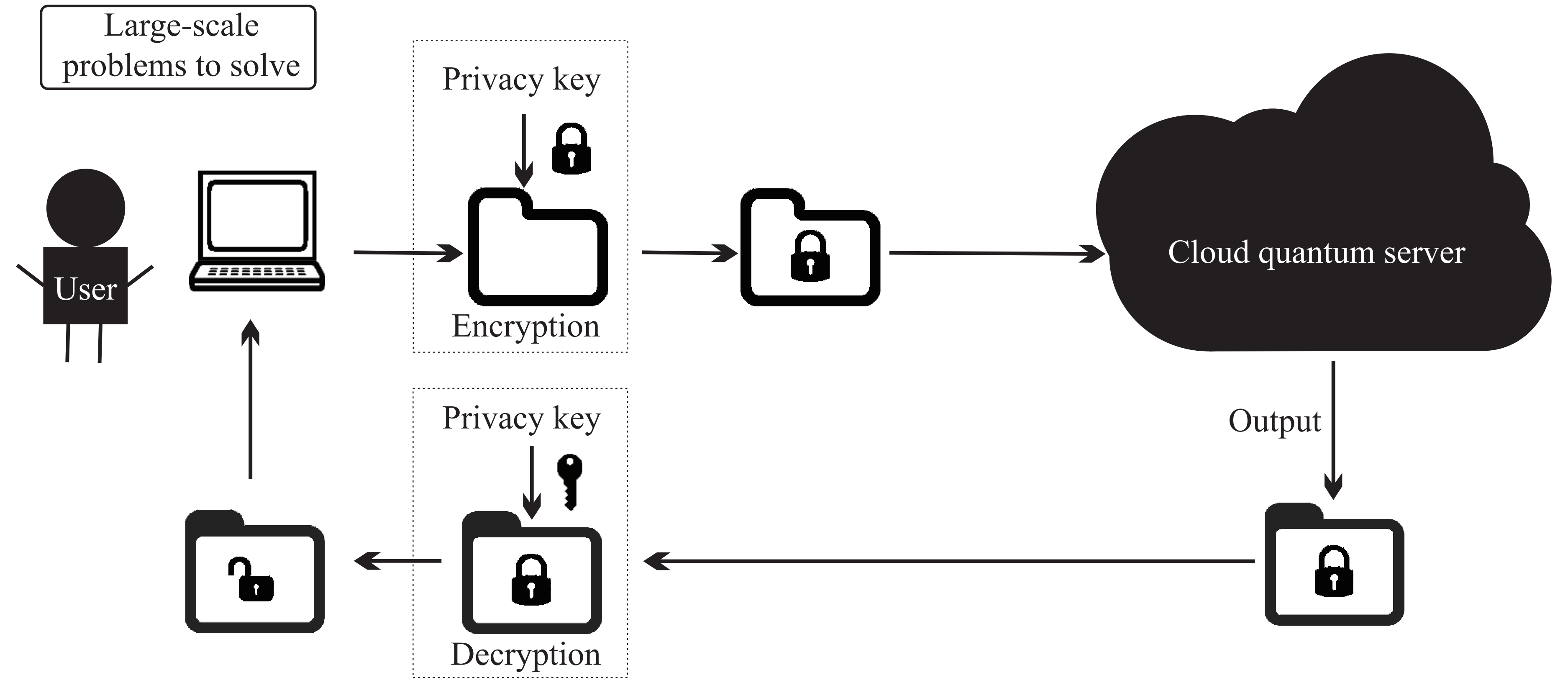}
\caption{The homomorphic encryption scheme for solving linear equations using cloud quantum computer. Taken from~\cite{huang2017homomorphic}.}
\label{fig:Figure 6.5}
\end{figure}

Quantum computers promise to solve complex problems that today's classical computers cannot solve, such as Shor's factoring algorithm ~\cite{shor1999polynomial}, and Harrow-Hassidim-Lloyd (HHL) related algorithms for big data processing ~\cite{harrow2009quantum,rebentrost2014quantum,rebentrost2018quantum,wiebe2012quantum,biamonte2017quantum}. In 2017, Zheng $et$ $al.$ used a four-qubit superconducting quantum processor to implement HHL algorithm for solving a two-dimensional system of linear equations ~\cite{zheng2017solving} (see Fig.~\ref{fig:Figure 6.4}). Huang $et$ $al.$ first applied homomorphic encryption on IBM's cloud quantum computer platform for solving linear equations while protecting our privacy ~\cite{huang2017homomorphic} (see Fig.~\ref{fig:Figure 6.5}). These works show the potential applications of quantum computing, but they are not practical on NISQ devices since they rely on the development of error correction.

\begin{figure}[!t]
\centering
\includegraphics[width=0.95\columnwidth]{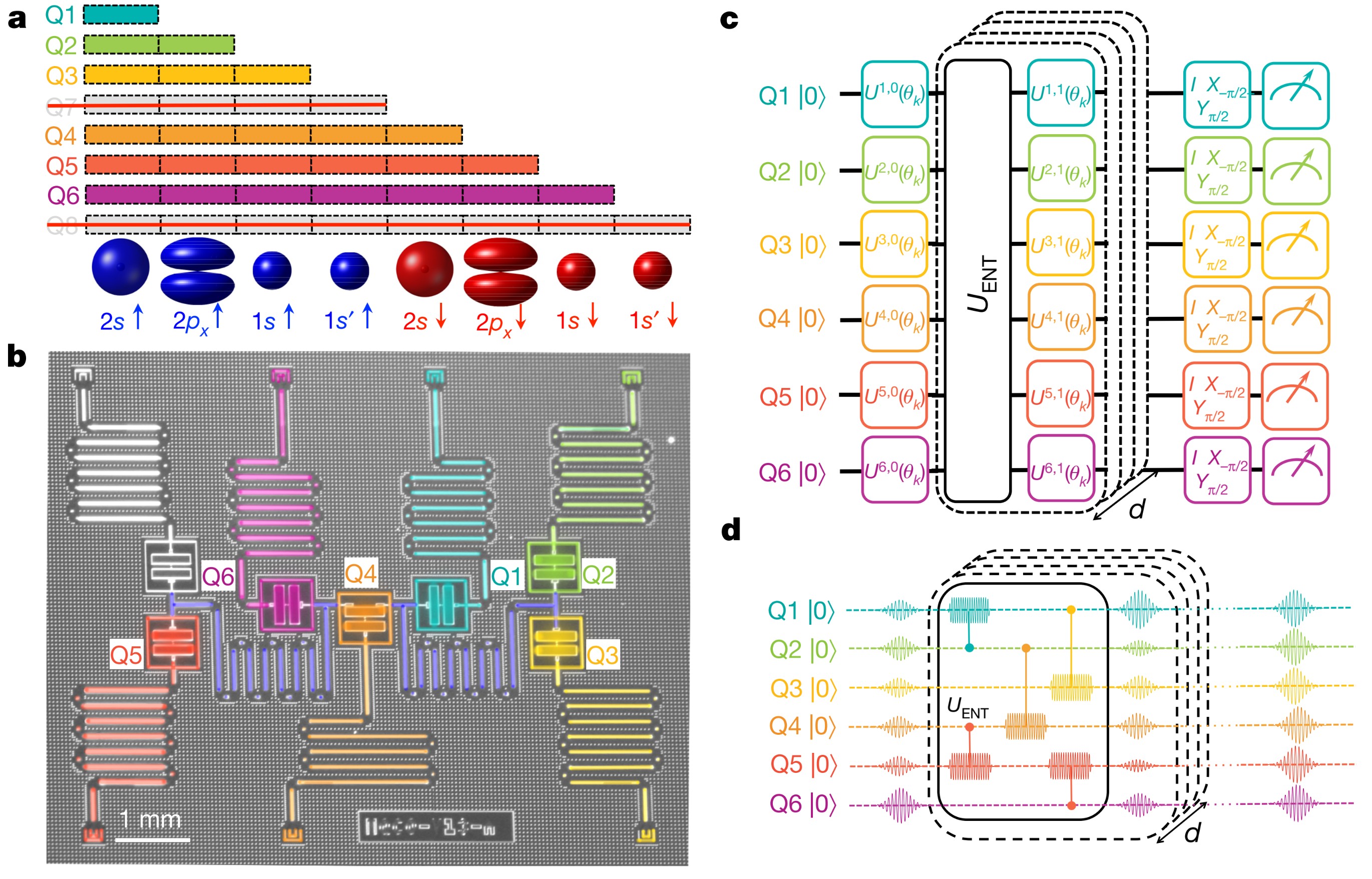}
\caption{Quantum chemistry on a superconducting quantum processor. (a) Parity mapping of spin orbitals to qubits. (b) Optical micrograph of the superconducting quantum processor with seven transmon qubits. (c) Quantum circuit for trial-state preparation and energy estimation. (d) An example of the pulse sequence for the preparation of a six-qubit trial state. Taken from~\cite{kandala2017hardware}.}
\label{fig:Figure 6.6}
\end{figure}

Hybrid quantum-classical algorithms with parameterized quantum circuits provide a promising approach for exploiting the potential of NISQ devices, and the variational quantum eigensolver (VQE) ~\cite{peruzzo2013variational,mcclean2016theory} and quantum machine learning ~\cite{benedetti2019parameterized,liu2019hybrid,lloyd2018quantum,schuld2019quantum} are the leading candidates for practical applications of NISQ devices. Colless $et$ $al.$ demonstrated a complete implementation of the VQE to calculate the complete energy spectrum of the $\rm H_2$ molecule with near chemical accuracy ~\cite{colless2017robust}. Kandala $et$ $al.$ demonstrated VQE on a six-qubit superconducting quantum processor, and showed the ability to addressing molecular problems beyond period I elements, up to $\rm BeH_2$ ~\cite{kandala2017hardware} (see Fig.~\ref{fig:Figure 6.6}). Chen $et$ $al.$ used an adiabatic variational hybrid algorithm, and demonstrated that many-body eigenstates can be efficiently prepared by an adiabatic variational algorithm assisted with a 3-qubit superconducting coprocessor~\cite{chen2019demonstration}.

\begin{figure}[!t]
\centering
\includegraphics[width=0.85\columnwidth]{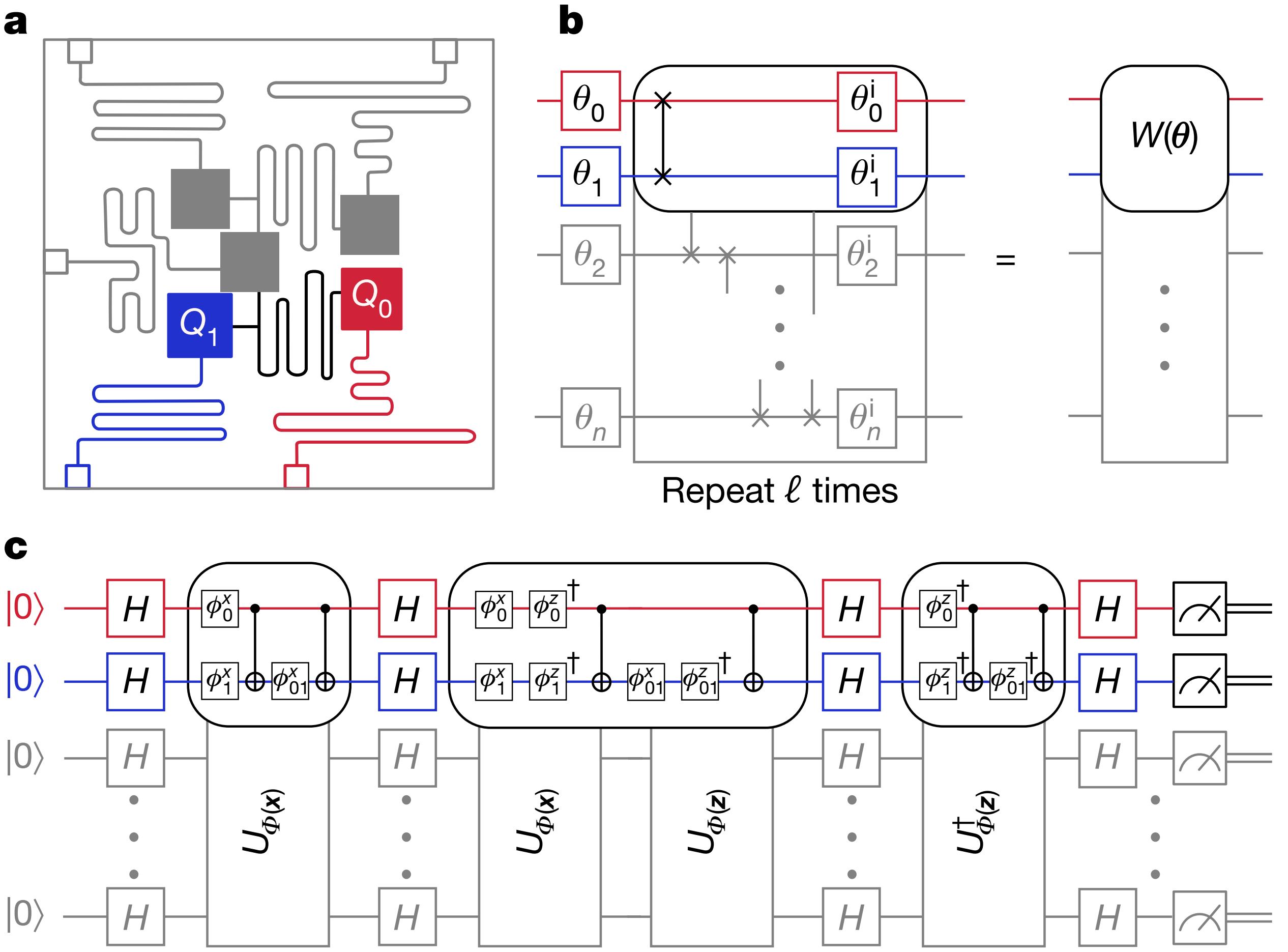}
\caption{Experimental implementations for supervised learning using superconducting quantum processor. (a) Quantum processor. (b) Variational circuit used for classification. (c) Circuit to directly estimate the fidelity between a pair of feature vectors. Taken from~\cite{havlivcek2019supervised}.}
\label{fig:Figure 6.7}
\end{figure}

\begin{figure}[!t]
\centering
\includegraphics[width=1\columnwidth]{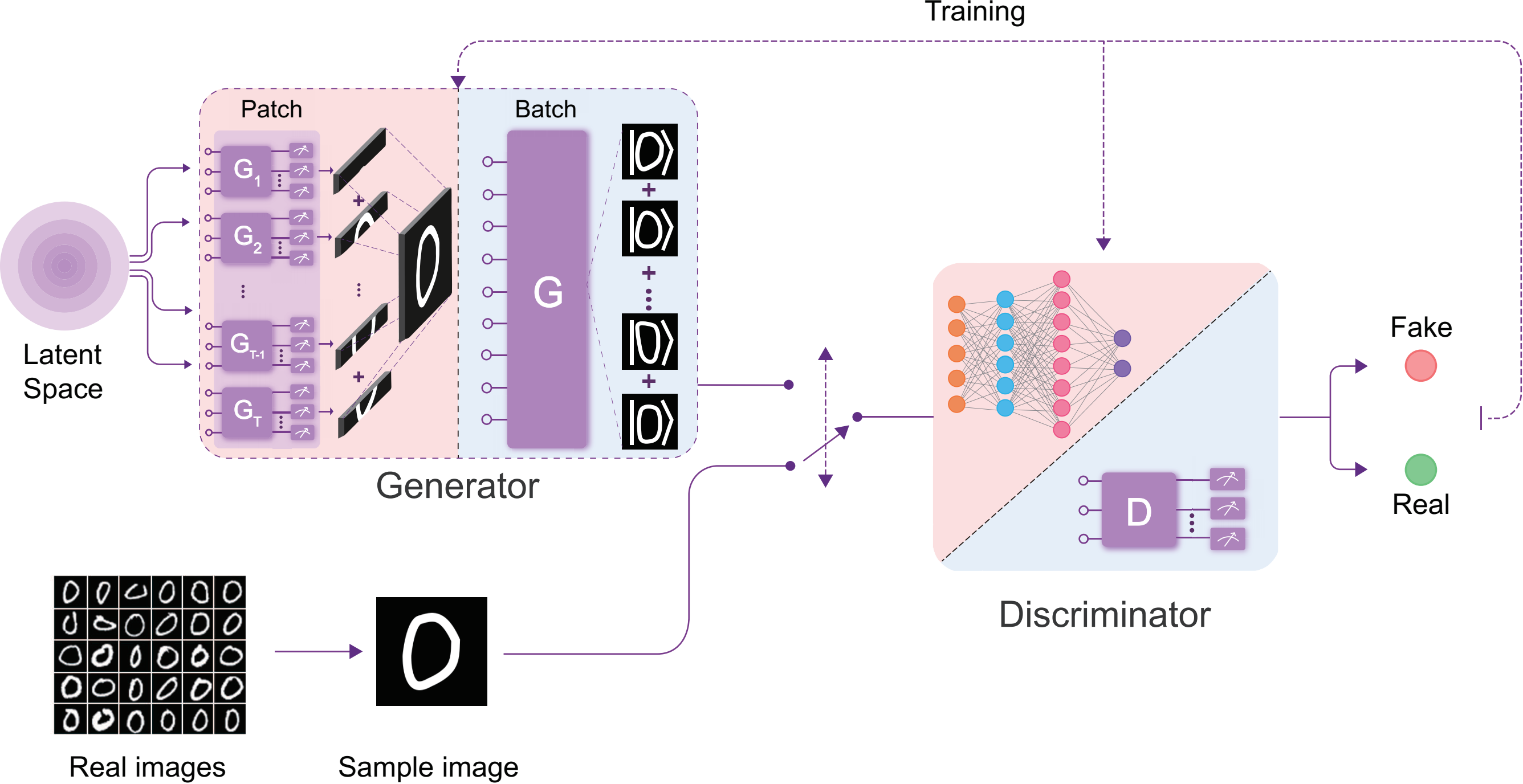}
\caption{The quantum generative adversarial networks scheme for hand-written digit images generation. Taken from~\cite{huang2020experimental}.}
\label{fig:new-QGAN}
\end{figure}

Quantum machine learning with parameterized quantum circuits is another hot topic for NISQ devices. Havl{\'\i}{\v{c}}ek $et$ $al.$ implemented two classifiers, a variational quantum classifier and a quantum kernel estimator, on a superconducting quantum processor ~\cite{havlivcek2019supervised} (see Fig.~\ref{fig:Figure 6.7}), which presents a route to accurate classification with NISQ hardware. Some quantum generative models for toy problems ~\cite{zoufal2019quantum,zhu2019training,hu2019quantumgenerative} were implemented on superconducting quantum processor. And recently, the first experiment to generate real-world hand-written digit images was achieved on a superconducting quantum processor (see Fig.~\ref{fig:new-QGAN})~\cite{huang2020experimental}, and the experimental results clearly show that quantum generative adversarial networks (GANs) can achieve comparable (or even better) performance to the classical GANs based on multilayer perceptron and convolutional neural network architectures, respectively, with similar number of parameters and similar training rounds. In the next stage, the implementation of quantum machine learning on near-term quantum devices to solve more complex real-world learning tasks with real advantage will become a key issue and interesting task in the field of quantum computing.

\subsection{Quantum supremacy}

\begin{figure}[!t]
\centering
\includegraphics[width=0.8\columnwidth]{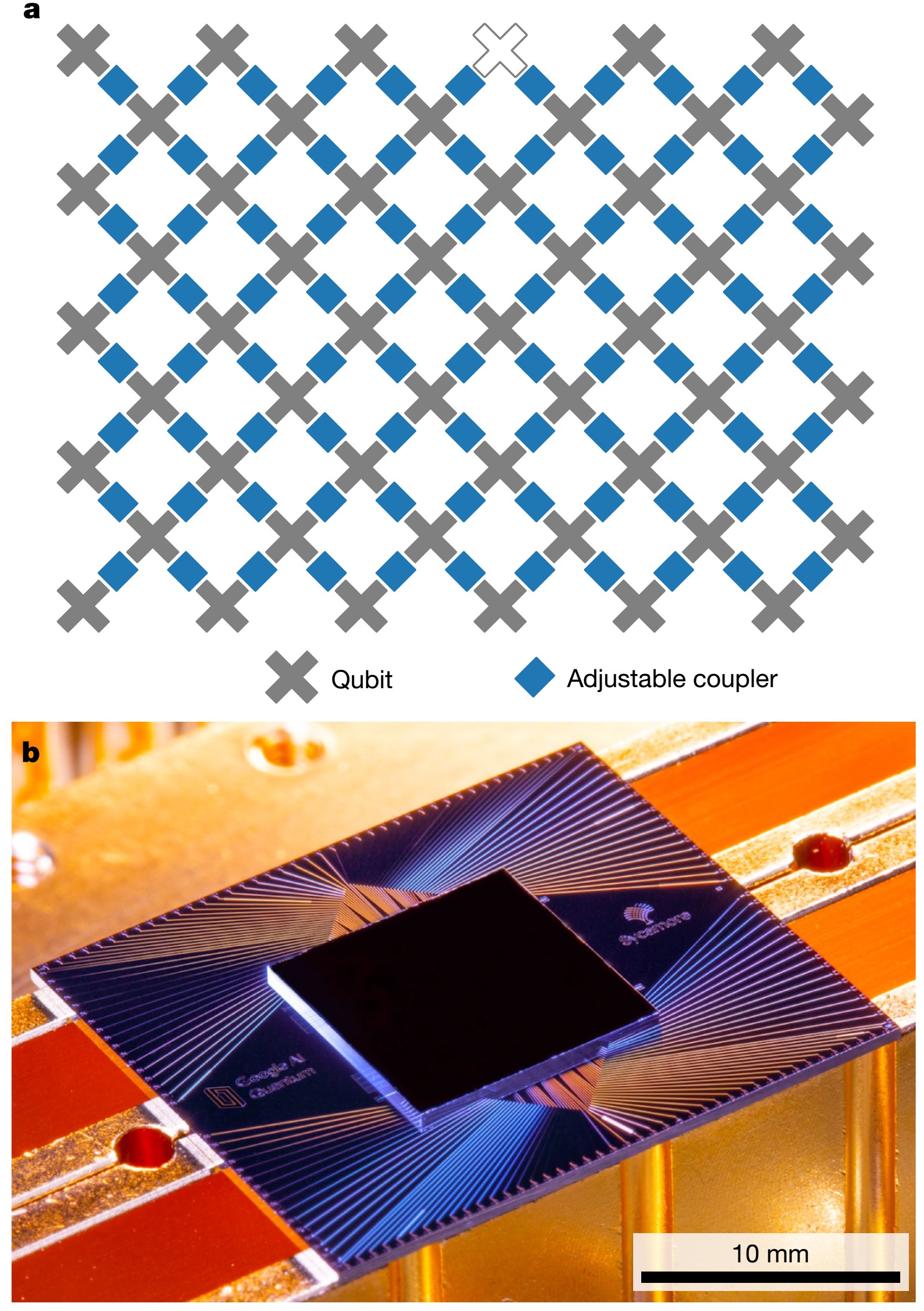}
\caption{The Sycamore processor. (a) Layout of processor. (b) Photograph of the Sycamore chip. Taken from~\cite{arute2019quantum}.}
\label{fig:Figure 6.8}
\end{figure}

Quantum supremacy is the goal of demonstrating that a programmable quantum device can solve a problem that classical computers practically cannot~\cite{harrow2017quantum,boixo2018characterizing,neill2018blueprint}. In 2019, Google officially announced that it achieved the milestone of ``quantum supremacy"~\cite{arute2019quantum}. They performed the random circuit sampling on a programmable superconducting quantum processor with 53 available qubits called Sycamore (see Fig.~\ref{fig:Figure 6.8}). In their work, the improvement of the 2-qubit gate fidelity, and the reduction of crosstalks in parallel gate operations are critical techniques for this progress. And finally, they showed a computational experiment that implements an impressively large two-qubit gate quantum circuit of depth 20, with 430 two-qubit and 1,113 single-qubit gates, and with predicted total fidelity of 0.2$\%$.

\section{Summary and Outlook}

In this review we have presented basic notions and recent advances in the field of superconducting quantum computing. Recent years have seen an increased interest in superconducting quantum computing, and we believe there is significant promise for the long term. With the advent of intermediate scale quantum devices, quantum computing will be used as a tool for computing tasks and gradually play an increasingly important role in science and engineering. Despite the great advances that have been achieved, a number of challenges and open questions we are facing in both theoretical and experimental work. There is still the need to develop and implement higher quality superconducting qubit, such as improving the qubit connectivity, the gate fidelity, and coherence time, which are the key challenge for the development of superconducting quantum computing. It is remain a question what exactly is the first ``killer app'' implemented by quantum computing in the future, and as the realization of quantum computing seems more and more certain, the pressure to find killer apps for quantum computing grows. We need to find a few killer applications which really do show a distinctive quantum advantage over classical computing for near term devices and also long term devices. The quantum supremacy achievement marks just the first of many steps necessary to develop practical quantum computers, and perhaps the next two big milestones are to actually do something useful on near-term devices, and develop an ``error correcting" quantum computer.

\begin{acknowledgments}
\textbf{Acknowledgments.} We thank Franco Nori, G{\"o}ran Wedin, and Anton Frisk Kockum for helpful discussions. This research was supported by the National Key Research and Development Program of China (grant nos. 2017YFA0304300), NSFC (grant nos. 11574380), the Chinese Academy of Science and its Strategic Priority Research Program (grant no. XDB28000000), the Science and Technology Committee of Shanghai Municipality, and Anhui Initiative in Quantum Information Technologies. H.-L. Huang acknowledges support from the Open Research Fund from State Key Laboratory of High Performance Computing of China (Grant No. 201901-01), National Natural Science Foundation of China under Grants No. 11905294, and China Postdoctoral Science Foundation.
\end{acknowledgments}

\bibliographystyle{naturemag}
\bibliography{SuperNotes}



\end{document}